\def\um {\,$\mu$m }
\def\mathum {\,\mu {\rm m}}
\def\pmcol {r @{$\,\pm\,$\extracolsep{0pt}} l @{\extracolsep{\fill}}}

\def\twocol {r @{\extracolsep{0pt}} l @{\extracolsep{\fill}}}
\def\cc {\multicolumn{2}{c}}
\def\nopmdata {\cc{\nodata}}

\def\hour {$^{\rm h}$}
\def\min {$^{\rm m}$}

\def\adeg {$^\circ$}
\def\fsec {$^{\rm s}$.}
\def\fasec {''.}

\documentclass[iop,apj,tighten]{emulateapj}
\usepackage{apjfonts}

\usepackage{latexsym}
\usepackage{graphics}
\usepackage{rotating}
\usepackage{hyperref}

\newcounter{subfigure}

\shorttitle{FIR Properties of {\em Spitzer}-selected Starbursts}
\shortauthors{Kov\'{a}cs et al.}

\begin{document}

\title{Far-Infrared Properties of {\em Spitzer}-selected Luminous Starbursts}

\author{
A.~Kov\'{a}cs\altaffilmark{1,2}, 
A.~Omont\altaffilmark{3,4}, 
A.~Beelen\altaffilmark{5}, 
C.~Lonsdale,\altaffilmark{6,7},
M.~Polletta\altaffilmark{7},
N.~Fiolet\altaffilmark{3,4},
T.~R.~Greve\altaffilmark{8,9},
C.~Borys\altaffilmark{10},
P.~Cox\altaffilmark{11},
C.~De~Breuck\altaffilmark{12},
H.~Dole\altaffilmark{5},
C.~D.~Dowell\altaffilmark{10}, 
D.~Farrah\altaffilmark{13},
G.~Lagache\altaffilmark{5},
K.~M.~Menten\altaffilmark{2},
T.~A.~Bell\altaffilmark{10}, 
F.~Owen\altaffilmark{14}
}

\affil{
$^1$University of Minnesota, 116 Church St SE, Minneapolis, MN 55414, USA; \\
$^2$Max-Plank-Institut f\"{u}r Radioastronomie, Auf dem H\"{u}gel 69, 53121 Bonn, Germany\\
$^3$UPMC Univ Paris 06, UMR7095, Institut d'Astrophysique de Paris, F-75014, Paris, France\\
$^4$CNRS, UMR7095, Institut d'Astrophysique de Paris, F-75014, Paris, France \\
$^5$Institut d'Astrophysique Spatiale, bat 121, Universit\'{e} Paris Sud 11 \& CNRS (UMR8617), 91405 Orsay Cedex, France \\
$^6$Infrared Processing \& Analysis Center, California Institute of Technology, 100-22, Pasadena, CA 91125, USA\\
$^7$Center for Astrophysics \& Space Sciences, University of California, San Diego, La Jolla, CA 92093-0424, USA\\
$^8$Max-Planck-Institut f\"{u}r Astronomie, 69117 Heidelberg, Germany\\
$^9$Dark Cosmology Centre, Niels Bohr Institute, University of Copenhagen, Juliane Maries Vej 30, DK-2100 Copenhagen \O, Denmark \\
$^{10}$California Institute of Technology, 1200 East California Boulevard, Pasadena, CA 91125, USA\\
$^{11}$Institut de Radioastronomie Millimetrique, 200 rue de la Piscine, 380406 St.\ Martin d'Heres, France\\
$^{12}$European Southern Observatory, Karl-Schwarzschild Strasse, 85748 Garching bei M\"{u}nchen, Germany\\
$^{13}$Department of Physics \& Astronomy, University of Sussex, Falmer, Brighton, BN1 9RH, UK\\
$^{14}$National Radio Astronomy Observatory, P.O. Box 0, Socorro, NM 87801, USA
}

\journalinfo{The Astrophysical Journal, 717:29-39, 2010 {\rm July} 1}
\submitted{Received 2009 December 21; Accepted 2010 April 5}

\email{kovacs @ astro.umn.edu}

\begin{abstract}
We present SHARC-2 350\um data on 20 luminous $z \sim 2$ starbursts with $S_{1.2{\rm mm}}$$>$2\,mJy from the {\em Spitzer}-selected samples of Lonsdale et al.\ and Fiolet et al. All the sources were detected, with $S_{350\mu{\rm m}}$$>$25\,mJy for 18 of them. With the data, we determine precise dust temperatures and luminosities for these galaxies using both single-temperature fits and models with power-law mass--temperature distributions. We derive appropriate formulae to use when optical depths are non-negligible. Our models provide an excellent fit to the 6$\mathum$--2\,mm measurements of local starbursts.
We find characteristic single-component temperatures $T_1$$\simeq$35.5$\pm$2.2\,K and integrated infrared (IR) luminosities around 10$^{12.9\pm0.1}$\,L$_{\odot}$ for the SWIRE-selected sources. Molecular gas masses are estimated at $\simeq$4$\times$10$^{10}$\,$M_{\odot}$, assuming $\kappa_{850\mu{\rm m}}$=0.15\,m$^2$\,kg$^{-1}$ and a submillimeter-selected galaxy (SMG)-like gas-to-dust mass ratio. The best-fit models imply $\gtrsim$2\,kpc emission scales. We also note a tight correlation between rest-frame 1.4\,GHz radio and IR luminosities confirming star formation as the predominant power source.
The far-IR properties of our sample are indistinguishable from the purely submillimeter-selected populations from current surveys. We therefore conclude that our original selection criteria, based on mid-IR colors and 24\um flux densities, provides an effective means for the study of SMGs at $z$$\sim$1.5--2.5.
\end{abstract}

\keywords{galaxies: high-redshift --- galaxies: ISM --- galaxies: photometry --- galaxies: starburst --- infrared: galaxies --- submillimeter: galaxies}

\maketitle

\section{Introduction}

Star-forming galaxies release a significant fraction of their energy output at infrared (IR) and submillimeter wavelengths ($\lambda$$\sim$5--1000\um in the rest frame). The luminosities of starbursts, with star formation rates over 100\,M$_{\odot}$\,yr$^{-1}$, are almost exclusively carried at IR wavelengths. Whereas luminous and ultra-luminous infrared galaxies (LIRGs and ULIRGs) are extremely rare in the local universe, they are prevalent at higher redshifts \citep[e.g.][]{LeFloch2005, Caputi2007}. By studying this population we can learn about the star formation history of the universe, and because the starbursting is triggered by merger events, we can also probe models of structure formation and halo dynamics through the ages.

Around 1\,mm wavelengths it is possible to find similar starbursts across much of the volume of the universe, because the dimming of radiation from increasing distances is countered by the steeply rising energy spectrum between 2\,mm and 200$\mathum$ in the rest frame. Thus, an IR galaxy would produce nearly the same (sub)millimeter flux regardless of its distance in the range of $z \sim 0.5$--10.

The nearly bias-free selection, together with the fundamental desire for studying starbursts, make submillimeter surveys \citep{Coppin2006, Weiss2009, Austermann2009, Ivison2009} especially relevant. However, while finding submillimeter-selected galaxies (SMGs) is relatively straightforward, studying them in any detail has proved difficult. This is due to two factors: they are faint at other wavelengths, and there are often multiple possible optical/NIR counterparts \citep[see][]{Pope2006, Younger2009b} due to the poor spatial resolution of most (sub)millimeter telescopes (typically 10''--30'' FWHM). 

As a result, most spectroscopic redshifts have been obtained for SMGs with radio-detected counterparts providing the required positional accuracy. Thus, \citet{Chapman2003, Chapman2005} and \citet{Kovacs2006} discovered that most SMGs lie around a median redshift of 2.3, are extraordinarily luminous (10$^{12}$--10$^{13}$\,L$_\odot$) and have large molecular gas reservoirs (10$^{10}$--10$^{11}$\,M$_\odot$). The close correlation between IR and radio luminosities of SMGs resembles that of local star-forming galaxies \citep{Helou1985, CondonBroderick1991, Condon1992, Yun2001}, implying that star-formation is the principal power source.

Despite these successes however, the number of SMGs with spectroscopic redshifts is only around a hundred, and less than half of these are characterized in the far-infrared (FIR). Furthermore, the radio-undetected SMG population is almost completely unexplored. A truly unbiased understanding of SMGs requires improved selection methods.

\subsection{SMGs and {\em Spitzer}}

A major step forward in understanding SMGs was provided by {\em Spitzer}. With its sensitive mid-IR imaging and spectroscopic capabilities, {\em Spitzer} detected most SMGs in multiple mid-IR bands, which can provide accurate photometric redshifts $dz/(1+z)$$\simeq$0.1 \citep[][Lonsdale et al.~2009]{Pope2006}. Furthermore, many {\em Spitzer}-selected objects are predicted to be submillimeter bright (Lonsdale et al.~2009; hereafter \citet{Lonsdale2009}). Unfortunately, the identification of mid-IR counterparts to SMGs is plagued by problems similar to those of optical association \citep{Pope2006}, and so the reliance on radio associations remains. 

Selecting strongly luminous starbursts at $z$$\sim$2 from {\em Spitzer} data can overcome these problems, and the resulting samples are expected to contain a large proportion of SMGs within them. Thus, a two-step process has been suggested. First, selecting sources with peak {\em Spitzer}--Infrared Array Camera (IRAC) flux densities in the 5.8\um channel should remove galaxies with a mid-IR luminous active galactic nucleus \citep[AGN;][]{Weedman2006, Farrah2008}, leaving objects with a clear rest-frame 1.6\um opacity minimum at $z$$\approx$2. The required IRAC detections ensure that stellar masses are large, especially when applied to limited sensitivity samples such as SWIRE \citep{Lonsdale2003}. Second, a bright 24\um flux density cut ($S_{24\mathum}$$>$400\,$\mu$Jy) favors starbursts at the same redshift as the strong rest-frame 7.7\um polycyclic aromatic hydrocarbon (PAH) emission feature is redshifted into the band. Redshifts from InfraRed Spectrograph (IRS) spectra or from IRAC photometry confirm that most sources selected in this way lie at $z$$\simeq$2 \citep[][N.\ Fiolet et al., in preparation]{Weedman2006}. The effectiveness of such a two-pronged selection was confirmed by \citet{Lonsdale2009} and Fiolet et al.~(2009, hereafter \citet{Fiolet2009}), who detected a significant fraction of these sources at 1.2\,mm using the MAMBO camera at the IRAM 30\,m telescope (i.e.,~$S_{1.2_{\rm mm}}$$\gtrsim$2\,mJy). Clearly, the proposed {\em Spitzer} selection (see Table~\ref{tab:selection}) yields distant luminous starburst galaxies without radio preselection. 

\begin{deluxetable}{l c}[!bth]
\tablewidth{\columnwidth}
\tablecolumns{2}
\tablecaption{Sample Selection Criteria\label{tab:selection}}
\tablehead{
  Band/Instrument & Criterion
}
\startdata
IRAC & $S_{3.6\mathum} < S_{4.5\mathum} < S_{5.8\mathum} > S_{8.0\mathum}$ \\[1pt]
MIPS 24 & $S_{24\mathum} > 400$\,$\mu$Jy \\[1pt]
MAMBO & $S_{1.2{\rm mm}} \gtrsim 2$\,mJy
\enddata
\tablecomments{Summary of the selection criteria for the MAMBO-detected 5.8$\mathum$ peaker samples of \citet{Lonsdale2009} and \citet{Fiolet2009}.
}
\end{deluxetable}

A key question is to what extent the SWIRE-selected samples are in fact representative of the purely submillimeter-selected population. A straightforward way to test this is to constrain where the peak of the IR emission lies, thus bridging the gap between the cold dust detected by MAMBO, and the hot dust and PAHs detected by {\em Spitzer}. Such constraints can be used to derive effective dust temperatures and accurate IR luminosities, for comparison with the overall SMG population. Because the sample consists of galaxies at z$\sim$2, and dust temperatures are expected in the range of 30--40\,K, the 350\um band offers an ideal opportunity for providing the constraints we seek. 

For our study, we targeted 12 galaxies from \citet{Fiolet2009} and another 8 from \citet{Lonsdale2009}. The former is an almost complete sample of sources from 0.5\,deg$^2$ satisfying the criteria of Table~\ref{tab:selection}, whereas the \citet{Lonsdale2009} sources are drawn arbitrarily from $\sim$30\,deg$^2$. To enhance our chances of detection at 350$\mathum$, we selected objects with clear MAMBO detections, with an additional preference toward the highest 1.2\,mm flux density measurements.

The observations are described in Section~\ref{sec:observations} and the results are summarized in Section~\ref{sec:results}, where we also discuss one of the submillimeter sources, which appears to be a close association of several IR galaxies. In Section~\ref{sec:sed}, we develop appropriate spectral energy distribution (SED) models to interpret the measurements. With these models, we characterize the sample in Section~\ref{sec:discussion}, deriving precise dust temperatures, dust/gas masses and luminosities, and with them we verify the (far-)infrared to radio correlation. 

We assume a cosmology with $H_0$=71\,km\,s$^{-1}$\,Mpc$^{-1}$, $\Omega_M$=0.27 and $\Omega_{\Lambda}$=0.73, and calculate distances as prescribed by \citet{Hogg1999}.

\section{SHARC-2 350\um Observations}
\label{sec:observations}

\begin{deluxetable}{@{\extracolsep{\fill}} l c r \pmcol @{\hspace{10pt}} l}[!bth]
\tablewidth{\columnwidth}
\tablecolumns{6}
\tablecaption{Variable Calibrators\label{tab:cal}}
\tablehead{
  \colhead{} & \colhead{} & \colhead{} & \cc{$S_{350\mathum}$} & \colhead{Cross} \\ 
  Name & \colhead{Date} & \colhead{$N_{\rm obs}$} & \cc{(Jy)} & \colhead{Calibrators}
}
\startdata
CIT6  &  2007 Jan 17    &  1 & 3.14&0.25 & Arp\,220 \\
      &  2007 Apr 17    &  2 & 3.12&0.11 & Arp\,220, Vesta \\
      &  2008 Jan 18    &  5 & 2.21&0.08 & Arp\,220 \\
      &  2009 Jan 22--25& 16 & 2.93&0.04 & Arp\,220, Ceres \\
\\[-3pt]
3C\,273 &  2007 Apr 19    &  1 & 4.08&0.30 & CIT6 \\
      &  2008 Feb 23    &  2 & 3.82&0.21 & Arp\,220 \\
      &  2008 Feb 24    &  2 & 3.84&0.20 & Arp\,220 \\
      &  2009 Jan 23    &  1 & 5.04&0.37 & Arp\,220 \\
      &  2009 Jan 24    &  3 & 5.46&0.22 & Arp\,220, Ceres \\
      &  2009 Jan 25    &  2 & 5.25&0.25 & Arp\,220, Ceres
\enddata
\tablecomments{The columns are: the name of the variable calibrator source, the period of measurements, the number of measurement, the derived 350\um flux densities, and the stable cross-calibrator sources used.}
\end{deluxetable}

350\um observations were carried out at the 10.4\,m Caltech Submillimeter Observatory (CSO), using the SHARC-2 Camera \citep{Dowell2003}. Data were collected during five successful observing runs between 2007 January and 2009 February.  The typical weather conditions during observing ranged from good to marginally useful, with in-band line-of-sight opacities $\tau$$\simeq$1--2, corresponding to precipitable water vapor levels around 0.7--1.3\,mm.

We took advantage of the open-loop actuated Dish Surface Optimization System \citep[DSOS;][]{Leong2006} of the telescope to maintain the optimal beam qualities at all elevations. As a result, we expect our telescope beam efficiencies to be relatively high and limited mainly by the 12\um surface rms of the CSO panels.

Our observing strategy was to modulate the telescope primary with small, field-of-view-sized open Lissajous patterns \citep{scanning, thesis}, with average scanning speeds around a beam per second. These patterns facilitate the separation of the astronomical source from the various correlated signals (of the atmosphere or the instrument). At the same time, the pattern provides uniform coverage over the field of view, the source remains exposed on the array at all times, and the scanning mode does not require extreme maneuverability of the telescope structure.

\begin{deluxetable*}{@{\extracolsep{\fill}} l l r @{,} l c \pmcol \pmcol \pmcol \pmcol \pmcol \pmcol}[!htbp]
\tablewidth{\textwidth}
\tablecaption{Summary of the FIR and Radio Observations\label{tab:flux densities}}
\tablecolumns{17}
\tablehead{ 
\colhead{} & \colhead{} & \cc{Offset} & \colhead{}
  & \cc{$S$(24$\mathum$)} & \cc{$S$(350$\mathum$)} & \cc{$S$(1.2\,mm)} & \cc{$S$(20\,cm)} & \cc{$S$(50\,cm)} & \cc{$S$(90\,cm)} \\
  ID & \colhead{IAU Name} & \cc{(arcsec)} & \colhead{$z$}
  & \cc{($\mu$Jy)} & \cc{(mJy)} & \cc{(mJy)} & \cc{($\mu$Jy)} & \cc{($\mu$Jy)} & \cc{($\mu$Jy)}
}
\startdata
LH-02 & SWIRE\_J103639.57+575346.6 & +2.2 & +1.0 & 1.93\tablenotemark{$a$}  & 838&15 & 54.8&5.8 & 5.29&0.85 & \nopmdata & \nopmdata & \nopmdata \\
LH-06 & SWIRE\_J103837.03+582214.7 & +3.2 &$-$0.0& 1.68\tablenotemark{$b$}  &1072&19 & 45.5&7.1& 3.83&0.84 & \nopmdata & \nopmdata & \nopmdata \\
LH-03 & SWIRE\_J104313.33+574621.0 & +1.5 &$-$0.2& 2.67\tablenotemark{$a$}  & 722&17 & 62.9&10.2& 3.79&0.76 & \nopmdata & \nopmdata & \nopmdata \\
L-1   & SWIRE\_J104351.16+590057.9 & +2.7 &$-$1.0& 2.26\tablenotemark{$c$}  & 722&17 & 34.1&6.7& 2.95&0.66 & 77&9.0 & 142&19 & 327&74 \\
L-9   & SWIRE\_J104440.25+585928.4 & +0.6 & +0.4 & 2.01\tablenotemark{$c$}  & 674&19 & 57.0&15.1& 4.00&0.55 & 117&9.2& 195&22 & 266&69 \\
L-11  & SWIRE\_J104556.90+585318.9 & +1.3 & +0.8 & 1.95\tablenotemark{$c$}  & 650&18 & 39.7&5.9 & 3.08&0.58 & 315&4.1& 425&43 & 651&72 \\
L-14  & SWIRE\_J104638.67+585612.6 & +1.6 &$-$0.1& 2.07\tablenotemark{$c$}  & 611&17 & 31.9&4.9 & 2.13&0.71 & 160&6.0& 322&33 & 426&70 \\
L-15  & SWIRE\_J104656.47+590235.5 & +6.3 &$-$0.9& 1.89\tablenotemark{$c$}  & 419&18 & 26.9&4.2 & 2.36&0.62 & 69&3.7 & 181&21 & 258&69 \\
L-17  & SWIRE\_J104704.97+592332.3 &$-$1.2& +0.4 & 1.99\tablenotemark{$c$}  & 647&18 & 39.6&8.5& 2.24&0.64 & 341&33   & 756&78 &1052&75 \\
L-20  & SWIRE\_J104717.96+590231.8 & +4.7 &$-$1.9& 2.10\tablenotemark{$c$}  & 617&19 & 49.7&6.5 & 2.66&0.78 &  51&4.7 & 166&19 & 264&72 \\
L-21  & SWIRE\_J104718.63+584318.1 & +0.3 & +5.3 & 1.78\tablenotemark{$d$}  & 447&16 & 41.4&7.4 & 3.09&0.81 & 138&36& 176&39 & 350&76  \\
L-22  & SWIRE\_J104720.49+591043.6 & +2.5 & +1.5 & 2.57\tablenotemark{$d$}  & 435&16 & 18.9&5.7 & 3.41&0.73 & 101&7.1 & 238&26 & 447&70 \\
L-23  & SWIRE\_J104726.44+585213.3 & +4   & $-$3 & 2.38\tablenotemark{$d$}  & 434&17 & 19  &9\tablenotemark{$e$}  & 3.13&0.86 & 86&7.2 & 172&21 & 242&71 \\
L-25  & SWIRE\_J104738.32+591010.0 & +1.2 &$-$0.3& 1.96\tablenotemark{$c$}  & 723&17 & 31.8&5.9 & 2.56&0.74 & 69&9.0 & 166&20 & 348&72 \\
L-27  & SWIRE\_J104744.59+591413.4 & +0.9 &$-$2.4& 2.20\tablenotemark{$c$}  & 523&18 & 28.7&6.5 & 2.48&0.73 & 77&14 &106&18 & \cc{$<$219} \\
LH-01 & SWIRE\_J105007.26+571651.0 & +1.8 & +2.3 & 1.35\tablenotemark{$a$}  & 756&12 & 46.6&7.6 & 5.65&0.74 & \nopmdata & \nopmdata & \nopmdata \\
EN1-01 & SWIRE\_J160343.08+551735.8 & +2.9 & +0.1 & 2.37\tablenotemark{$a$} & 699&14 & 42.7&8.2& 2.81&0.78 & \nopmdata & \nopmdata & \nopmdata \\
EN1-02 & SWIRE\_J160440.45+543103.0 &$-$0.5&$-$7.3& 1.93\tablenotemark{$a$} & 772&18 & 44.3&9.1& 2.46&0.81 & \nopmdata & \nopmdata & \nopmdata \\
EN1-04\tablenotemark{$f$} & SWIRE\_J161658.14+535319.3 & +0.2 & +0.6 & 2.00\tablenotemark{$a$}  & 657&15 & 27.6&10.2& 2.82&0.99 & \nopmdata & \nopmdata & \nopmdata \\
EN2-01 & SWIRE\_J163734.44+415151.4 & +0.9 & +3.8 & 1.85\tablenotemark{$c$} & 814&20 & 30.4&5.7 & 2.52&0.70 & \nopmdata & \nopmdata & \nopmdata
\enddata
\tablenotetext{$a$}{Photometric redshift from \citet{Lonsdale2009}.}
\tablenotetext{$b$}{IRS Spectroscopic redshift from \citet{Lonsdale2009}.}
\tablenotetext{$c$}{IRS Spectroscopic redshift from N.\ Fiolet et al.\ (in preparation).}
\tablenotetext{$d$}{Photometric redshift from \citet{Fiolet2009}.}
\tablenotetext{$e$}{The 350\um flux density of L-23 is from the undeconvolved image close to the instrument resolution (middle panel of Figure~\ref{fig:L-23}).}
\tablenotetext{$f$}{SWIRE\_J161658.14+535319.3 is the only source in our sample with a MIPS 160\um detection of 81$\pm$4\,mJy. However, this datum is not included in our fits.}
\tablecomments{Source IDs prefixed with LH or EN are from \citet{Lonsdale2009}, whereas those lettered with L are from \citet{Fiolet2009}. SHARC-2 detection offsets are with respect to the SWIRE catalog coordinates. 1.2\,mm flux densities are from \citet{Lonsdale2009} and \citet{Fiolet2009}, 20\,cm radio data from \citet{Owen2008}, 50\,cm value from F.\ Owen et al.\ (in preparation), while 90\,cm points are from \citep{Owen2009}. The quoted rms uncertainties do not include the uncertainties in the calibration.}
\end{deluxetable*}

In total, we spent 37.5 hr on the 20 objects observed. We aimed for 5\,$\sigma$ detections, or limiting rms noise levels of 10--12\,mJy\,beam$^{-1}$. Pointing and flux density calibration measurements were taken frequently, at least once per hour during the observing period. Our principal flux density calibrators were Mars, Arp\,220 (9.93$\pm$0.11\,Jy, based on our own calibration against planets and asteroids), and the minor planets Ceres, Juno and Vesta, for which appropriate temperature models have been derived by C.~D.~Dowell and are available through the SHARC-2 Web site.\footnote{\url{www.submm.caltech.edu/~sharc}} In addition, we relied on the evolved star CIT6, and the quasar 3C\,273 as additional calibrators, owing to their relative proximity to the targeted fields in the Lockman Hole. To counter the long-term variability of CIT6 we measured it against our main calibrator sources during each observing run, while 3C\,273 we cross-calibrated on a daily basis (see Table~\ref{tab:cal}). Based on the residual scatter of calibrator flux densities, we expect a 350\um calibration accuracy around 8\% for our science targets.

We checked and adjusted the focus settings several times during each night, more frequently near sunrise and sunset when variations tend to be greatest. To avoid the unnecessary smearing of our sources due to systematic pointing errors, the pointing data of each run were carefully analyzed for trends. Corrections were applied as required at the time of reduction. The final pointing accuracy of the SHARC-2 data is expected at $\simeq$3'' rms.

We reduced the data using the {\em CRUSH} software \citep[1.99-a5;][]{CRUSH, thesis}, which estimates and removes the correlated atmospheric and instrumental signals, solves for the relative detector gains, and determines the appropriate noise weighting of the time streams in an iterated pipeline scheme. To maintain calibration accuracy, the inherent filtering of point-source signals by the reduction steps is carefully estimated and corrections are applied internally by {\em CRUSH}.

\section{Results}
\label{sec:results}

We detected all 20 of the targeted sources. In a few cases, the detections are tentative. The measured 350\um flux densities are listed in Table~\ref{tab:flux densities}, along with all prior measurements relevant to this work. We reached rms sensitivities of 4--10\,mJy\,beam$^{-1}$ in all but one field. 80\% of the 350\um detections are robust ($>$4\,$\sigma$), two are 3--4\,$\sigma$ and another two are marginal ($<$3\,$\sigma$). However, even the lowest signal-to-noise detection is significant (with a confidence level $>$95\%) given the small, beam-sized search areas surrounding the IRAC (and nearby radio) positions. For the most part, the SHARC-2 positions are in excellent agreement with those from IRAC, with deviations at $\lesssim$3'' rms. 

\subsection{Submillimeter Multiplets}
\label{sec:multiplets}

\begin{figure*}
\centering
\includegraphics[width=0.8\textwidth]{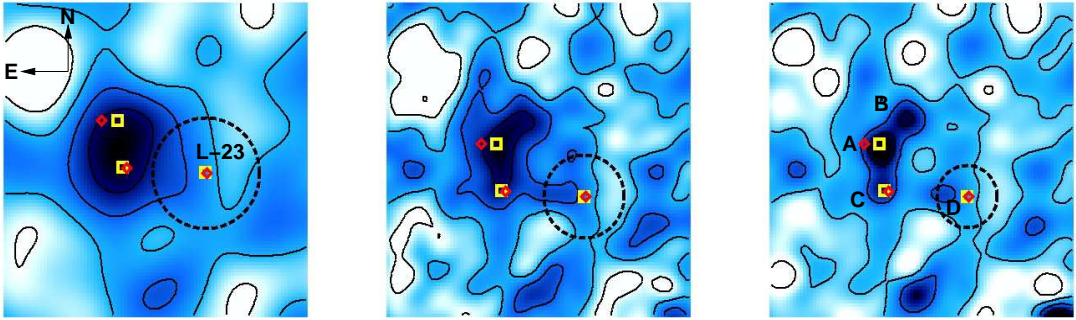}
\caption{45''$\times$45'' postage stamps around L-23. Left: the beam-smoothed image used for source extraction at $\sim$11'' resolution shows a single peak east of the expected SWIRE source position (dashed circle indicating typical search area). Center: near the diffraction-limited resolution of SHARC-2, the peak resolves into a banana-shaped feature and a hint of a point source close to the IRAC position. The image has been slightly realigned to match the elongated feature with two nearby IRAC galaxies (yellow squares). The nearby radio sources are shown with red diamonds. The search area is reduced as a result of the realignment. Right: a deconvolution suggests up to four distinct components (lettered A through D), all within the typical beam areas of current submillimeter surveys. Three of the peaks align well with IRAC sources in the field. Contours are -15, 0, 15, 30, and 45 mJy\,beam$^{-1}$ for the panels on the left and center, and -7.5, 0, 7.5, 15, and 22.5 mJy\,beam$^{-1}$ for the image on the right.
} 
\label{fig:L-23}
\end{figure*}

There is, however, the curious case of L-23 (see Figure~\ref{fig:L-23}), with an elongated $5.7$\,$\sigma$ SHARC-2 peak 13''.5 from the IRAC and radio positions. The discrepancy is too large ($>$4\,$\sigma$) to be a statistical pointing error. Coincidentally, we have a large number of pointing measurements both before and after observing the galaxy in question, all reinforcing an absolute 350\um astrometric accuracy better than 3'' rms. Thus, the single SHARC-2 peak is inconsistent with a source at the IRAC and radio positions. 

A substructure of the peak is revealed if we smooth less than the optimal amount for point source extraction. We find a banana-shaped feature dominating the 350\um emission to the east, and a fainter $\simeq$2\,$\sigma$ compact peak close to the expected position. The position and point-like nature of the fainter component is consistent with both our pointing accuracy, and the expected size ($\lesssim$1'') of a galaxy at $z$=2.38. Therefore, we identify this as the 350\um counterpart to L-23.

\begin{deluxetable}{l c c c l}[!htb]
\tablewidth{\columnwidth}
\tablecolumns{5}
\tablecaption{350\um Decomposition of L-23\label{tab:L-23}}
\tablehead{
  ID & \colhead{$\alpha_{2000}$} & \colhead{$\delta_{2000}$} & \colhead{$S_{350\mathum}$} & \colhead{Association}
}
\startdata
A  &  10\hour47\min28\fsec1 & +58\adeg52'20'' & 25\,mJy &  IRAC/MIPS24, radio \\
B  &  10\hour47\min27\fsec6 & +58\adeg52'24'' & 17\,mJy &  Unknown \\
C  &  10\hour47\min28\fsec1 & +58\adeg52'15'' & 14\,mJy &  IRAC/MIPS24, radio \\
D  &  10\hour47\min26\fsec9 & +58\adeg52'14'' & 10\,mJy &  L-23
\enddata
\tablecomments{The most likely decomposition of the blended 350\um emission from the neighborhood of L-23 \citep{Fiolet2009}, ordered in decreasing estimated brightness. The SWIRE/MAMBO source we identify as component D. The uncertainties in the deconvolved flux densities we estimate at around 5\,mJy. Positions are estimated to be accurate to $\lesssim$2'' rms. The positions of the associated IRAC/MIPS and radio sources are discussed in Section~\ref{sec:multiplets}.
}
\end{deluxetable}

Deconvolution of the elongated feature suggests three distinct components. Two of these align well with nearby IRAC and MIPS 24\um peaks (at 10\hour47\min28\fsec14, $+$58\adeg52'20\fasec9 and 10\hour47\min28\fsec04, $+$58\adeg52'14\fasec1) and radio sources (at 10\hour47\min28\fsec45, $+$58\adeg52'20\fasec9 and 10\hour47\min27\fsec97, $+$58\adeg52'14\fasec1), while the third peak is perhaps a super-resolution artifact. The positions and approximate brightness of the deconvolved peaks are summarized in Table~\ref{tab:L-23}. 

It is not unusual to find resolved pairs of submillimeter sources with $\sim$20''--30'' separation \citep{Blain2004, Greve2004, Farrah2006, Weiss2009}. Such pairs are relatively infrequent and correspond to massive dark matter halos ($>$10$^{13}$\,M$_\odot$). However, the clustering of SMGs on smaller scales is not well constrained. Therefore, the blending of as many as three or four galaxies into a single 350\um detection peak (with total $S_{350\mathum}$$\approx$60\,mJy) is interesting. 

We often assume that each deep-field (sub)millimeter detection corresponds to just one luminous galaxy, with unique counterparts at optical, IR, or radio wavelength. Granted that the clustering of SMGs on scales $\lesssim$15'' is not well known, it seems reasonable given the generally low density ($\sim$400/deg$^2$) of sources in the current, shallow (sub)millimeter surveys. However, our multiplet puts this assumption to the test: the 350\um components that we identify around L-23 would constitute a single source under the spatial resolution (15''--30'') of current (sub)millimeter surveys. The fact that it is a triplet or a quadruplet suggests that unresolved doublet, or even triplet (sub)millimeter sources may abound. The $\sim$20\% fraction of the radio-identified SMGs with multiple radio counterparts \citep{Ivison2007} could be an indication of the frequency of unresolved (sub)millimeter multiplets, which is also consistent with \citet{Aravena2010}. As such, the ubiquitous presumption that (sub)millimeter sources are individual galaxies may not be entirely justified. 

If multiple luminous galaxies are often behind the low-resolution (sub)millimeter peaks, the implications would be wide ranging. Some or all of the IRAC sources commonly detected close to an uncertain (sub)millimeter position \citep[e.g.][]{Pope2006, Younger2009b} could each contribute comparably to the measured (sub)millimeter flux densities. A commonality of multiplets would imply that (sub)millimeter source counts may be steeper than what SHADES \citep{Coppin2006}, AzTEC \citep{Austermann2009}, or the LABOCA deep-field survey \citep{Weiss2009} found; by counting multiple faint sources as a single bright one, these surveys would slightly underestimate the abundance of faint galaxies while substantially overestimating the number of bright ones (i.e., ULIRGs and hyper-LIRGs). Submillimeter surveys by {\em Herschel} (especially at 250/500$\mathum$) would be similarly affected. A strong clustering of SMGs at separation angles $\lesssim$20'', i.e., below the smallest scales that current (sub)millimeter surveys can resolve, would be implied. Unresolved multiplets could partially explain why as many as half of the SMGs in the literature have no identifiable radio counterparts: as opposed to having one detectable counterpart, there may be a number of associated faint radio sources below the detection level.





\section{Spectral Energy Distributions}
\label{sec:sed}

The addition of the 350\um data to the existing measurements provides a powerful constraint on the FIR SEDs of the SWIRE-selected samples of \citet{Lonsdale2009} and \citet{Fiolet2009}. It allows, for the first time, an accurate characterization of the dust temperatures and luminosities of individual sources in the sample, provided that physically realistic models of emission are fitted to the observations.

Before the emergence of details on the Wien-side shape of SEDs (e.g., from {\em Spitzer}), it was common practice to rely on single-temperature dust models. Extra temperature components were added as necessary to fit data at wavelengths $\lesssim$100\um. However, such models were clearly very na\"{i}ve. Among the many more physically motivated treatments is the pioneering work by \citet{DBP90} who considered dusty environments exposed to a range of radiation fields. Their framework has been fine-tuned to better match observations, especially of the aromatic features in the mid-IR \citep[e.g.][]{Dale2001, DaleHelou2002}. These models, fitted to well-studied local galaxies, serve as typical templates for interpreting {\em Spitzer} data.

Our goal is to derive accurate and robust FIR characterizations for our sample of galaxies. In so doing, we wish to move beyond a hodge-podge of temperature components but we also do not need the full complexity of the models with mid-IR detail. Thus, we fitted both single-temperature models (which offer direct comparisons to many previous studies) and power-law temperature distributions for a more accurate description. In the following sections, we shall discuss the models and the formulae we use for deriving the essential quantities.

\subsection{Single-temperature Dust Models}

The simplest model of the thermal IR emission assumes that it can be described as a singular gray-body at a physical temperature $T$ in the rest frame. Such single-$T$ flux densities $S_1$ are best expressed in terms of the observables: the measurement frequency $\nu_{\rm obs}$, and observed-frame temperature $T_{\rm obs} = T/(1+z)$, as

\begin{equation}
S_1(\nu_{\rm obs}, T_{\rm obs}) = m~d\Omega~ \left( 1-e^{-\tau} \right)~ B_{\nu_{\rm obs}}(T_{\rm obs}).
\label{eq:S1}
\end{equation}

The other terms are the physical solid angle $d\Omega$, the optical depth $\tau$, Planck's blackbody law $B_\nu$, and magnification $m$ (for gravitationally lensed objects). The physical solid angle is related to the projected source area $A$ by the angular-size distance $D_A$ as $d\Omega = A/D_A^2$. The optical depth can be estimated as the ratio of the total emitting cross-section of dust particles to the projected area of emission. As such, optical depths must be evaluated in the rest frame of the emission. Adopting a typical power-law frequency dependence for the emissivity for dust particles, optical depths can be expressed in the rest frame of the galaxy as

\begin{equation}
\tau(\nu_r) = \kappa_0~ \left[ \frac{\nu_r}{\nu_0} \right]^\beta \frac {M_d} {d\Omega D_A^2}.
\label{eq:tau}
\end{equation}

Here, $\nu_r = (1+z) \nu_{\rm obs}$ is the rest-frame frequency, $M_d$ is the total dust mass at an angular-size distance of $D_A$, $\kappa_0$ is the characteristic photon cross-section to mass ratio of particles at frequency $\nu_0$. The spectral emissivity index $\beta$ can be related to the fractal dimension $D$ of the dust particles as $\beta = D-1$ \citep{Yang2007}. Thus, $\beta$ values are expected in the range of 1--2, depending on the shape of particles. There have been several attempts for determining appropriate $\kappa$ values for interstellar dust particles \citep[e.g.~][]{Hildebrand1983, Krugel1990, Sodroski1997, James2002}. The resulting scatter of values reflects the difficulty of making such measurements. In this work, we assume $\kappa_{850\,\mu{\rm m}}$=0.15\,m$^2$\,kg$^{-1}$, which matches the value used by \citet{Kovacs2006} and is based on an average 125\um value \citep{Dunne2003} assuming $\beta$=1.5. As the strength of emission depends on the $\kappa_\nu M_d$ product, the inherent uncertainties in $\kappa$ ultimately affect the accuracy to which dust masses can be determined. At the same time, the temperatures and luminosities can be determined accurately and independently from the assumptions on the absorption efficiency.

We can calculate single-temperature IR luminosities $L_1$ by integrating the de-magnified observed flux density at luminosity distance $D_L$$=$$D_A (1+z)^2$ from the source:

\[
L_1(T) = 4 \pi~D_L^2~m^{-1} \int_0^\infty S_1(\nu, T) ~ d\nu
\]

The integration can be performed analytically in the optically thin limit ($\tau$$\ll$1), giving

\begin{equation}
L_1(T) = \Gamma(4+\beta) \zeta(4+\beta)~\frac {8 \pi h} {c^2} ~
 \left[ \frac{kT} {h} \right]^4 ~A_{\rm eff},
\label{eq:L1}
\end{equation}

in terms of the effective surface area of emission $A_{\rm eff}$, Planck's constant $h$, the Boltzmann constant $k$, the speed of light $c$, and Riemann's $\Gamma$ and $\zeta$ functions. The same expression is also applicable for the case of optically thick emission ($\tau$$\gg$1) by setting $\beta$=0. The effective area of emission $A_{\rm eff}$, in the optically thin and thick limits respectively, is given by

\begin{equation}
A_{\rm eff} \approx \left\{ \begin{array}{cl}
M_d \kappa_0  ( kT / h \nu_0 )^\beta & ~~~{\rm if}~\tau \ll 1
\\
\\ 
d\Omega D_A^2 & ~~~{\rm if}~\tau \gg 1~{\rm or}~ \beta \rightarrow 0
\end{array} \right.
\end{equation}

Given the relation of the projected surface area to total emitting area ($\pi r^2$ versus $4 \pi r^2$ for a sphere), our expression for the optically thick limit recovers the Stefan--Boltzmann law of blackbody luminosities. In the optically thin limit, $A_{\rm eff}$ is the physical emitting area at the frequency $\nu_r$=$kT/h$. From a computational point of view, it is practical to approximate the Riemann $\Gamma$ and $\zeta$ functions in Equation (\ref{eq:L1}) for the range of typical $\beta$ values:

\begin{equation}
\Gamma(4+\beta) \zeta(4+\beta) \approx 6.45 \exp(1.24\beta + 0.11\beta^2),
\end{equation}

which is accurate to about a percent in the range $\beta$$\simeq$0--2,

Interstellar dust emission is mostly optically thin at FIR wavelengths ($\lambda$$>$100$\mathum$). Only the more extreme galaxies and quasars, or strongly lensed objects, concentrate dust to such an extent that optical depths can become significant over the bulk of the emission spectrum. This is a concern, given the extreme nature of our objects. We therefore would like to also consider the effect of intermediate optical depths in deriving accurate luminosity measures.

\subsubsection{Intermediate Optical Depths}

\begin{deluxetable*}{@{\extracolsep{0pt}} l \twocol r r c \pmcol c c c c c c}[!tb]
\tablewidth{\textwidth}
\tablecolumns{16}
\tablecaption{FIR and Radio Properties of Local Starbursts\label{tab:localSB}}
\tablehead{
  \colhead{} & \cc{$D_L$} & \colhead{} &  \colhead{} & \colhead{} & \cc{$T_c$}  & \colhead{} & \colhead{$\log M_d$} & \colhead{$\log L$} & \colhead{} & \colhead{} & \colhead{} \\ 
  Name & \cc{(Mpc)} & \colhead{$N_{\rm IR}$} & \colhead{$N_{\rm rad}$} & \colhead{$|\chi_r|$} & \cc{(K)} &  \colhead{$\gamma$} & \colhead{($M_\odot$)}  & \colhead{($L_\odot$)} & \colhead{$\tau_{\rm pk}$} & \colhead{$q$} & \colhead{$\alpha$}
}
\startdata
Arp\,220          &  77&.6& 15 & 7 & 0.92 & 36.4&1.2 & 7.7(2) & 7.92(3) & 12.16(2) & 0.68 & 2.47(4)~ & 0.25(4) \\
IRAS\,20551-4250  & 188& &  10 & 1 & 0.99 & 37.9&2.4 & 7.2(3) & 7.33(9) & 12.09(3) & 0.19 & 2.61(9)~ & \nodata \\
IRAS\,22491-1808  & 348& &  14 & 1 & 0.68 & 39.0&2.0 & 7.6(3) & 7.62(8) & 12.20(3) & 0.38 & 2.91(7)~ & \nodata \\
M\,82             & 3&.57&  11 & 9 & 0.99 & 35.4&2.0 & 7.1(3) & 6.12(7) & 10.86(2) & 0.12 & 2.39(8)~ & 0.65(5) \\
NGC\,6090         & 127& &  11 & 1 & 1.81 & 28.0&1.9 & 6.9(3) & 7.44(8) & 11.67(2) & 0.15 & 2.33(12) & \nodata \\
NGC\,6240         & 105& &  21 & 7 & 1.07 & 32.3&1.1 & 7.1(1) & 7.56(4) & 11.96(2) & 0.25 & 1.84(4)~ & 0.85(3)
\enddata
\tablecomments{The FIR and radio properties of local starbursts, from the multi-$T$ fits to a selection of available IR continuum data in the 6$\mathum$--2\,mm range, with $\beta$=1.5 and assuming typical emission diameter of 1\,kpc (or 300\,pc for M\,82). The synchrotron component is modeled as a simple power-law $S_{\rm rad}$$\propto$$\nu^{-\alpha}$. Columns are: the object name, luminosity distance $D_L$, the number of IR and radio flux density data ($N_{\rm IR}$ and $N_{\rm rad}$) used, the residual scatter $|\chi_r|$ around the fit, the temperature $T_c$ of the dominant cold component, the mass--temperature index $\gamma$, dust masses $M_d$, integrated IR luminosities $L$, the optical depth $\tau$ around the IR emission peak $\nu_{\rm peak}$, the radio--(F)IR correlation constant $q$, and synchrotron spectral index $\alpha$. Uncertainties are 1\,$\sigma$ total errors. These are indicated in brackets for $\gamma$, $\log M_d$, $\log L$, $q$, and $\alpha$, and refer to the last digits in the quoted values. Dust masses assume $\kappa(850\mathum)$=0.15\,m$^2$\,kg$^{-1}$. Plots of these models and the data used for deriving them are available in the online edition.}
\label{tab:SB}
\end{deluxetable*}

We can derive approximations for the integrated luminosity for the general case of non-negligible optical depths. Our approach is based on the observation that most of the luminosity is radiated at frequencies close to the gray-body SED peak (see Equation (\ref{eq:S1})), which is observed near

\begin{equation}
\nu_{\rm pk} \approx (3+\beta)~ \frac{kT_{\rm obs}}{h}.
\label{eq:peak}
\end{equation}

Around the peak, we can approximate the frequency dependence of the emissivity, i.e.~$\epsilon = ( 1 - e^{-\tau} )$, with a power law $\epsilon \approx \epsilon_{\rm pk} (\nu/\nu_{\rm pk})^{\beta_{\rm eff}}$, where, by definition, an effective spectral index $\beta_{\rm eff}$ can be calculated as

\begin{equation}
\beta_{\rm eff} = \frac {\partial \log \epsilon} { \partial \log \nu } = \frac {\nu}{\epsilon} ~ \frac {\partial \epsilon}{\partial \nu}.
\end{equation}

Substituting the optical depth from Equation (\ref{eq:tau}) and evaluating at the peak $\tau_{\rm pk}$, we arrive at
 
\begin{equation}
\beta_{\rm eff} =  \beta~\frac { \tau_{\rm pk} } { e^{\tau_{\rm pk}} - 1 }.
\label{eq:beff}
\end{equation}

Thus, the frequency dependence of the emissivity near the peak of the emission can be characterized by an effective gray-body index $\beta_{\rm eff}$. This means that we can rely on Equation~(\ref{eq:L1}) to provide an approximation to the luminosity function, simply by replacing $\beta$ with $\beta_{\rm eff}$ above together with an effective emission area:

\begin{equation}
A_{\rm eff} \approx \frac { \left(1 - e^{ -\tau_{\rm pk}} \right) ~ d\Omega ~ D_A^2 }
{ (3+\beta)^{\beta_{\rm eff}} }.
\label{eq:Aeff2}
\end{equation}

It is possible to refine this approximation by seeking simultaneous solutions to Equations (\ref{eq:peak}) and (\ref{eq:beff}) with $\beta$$\rightarrow$$\beta_{\rm eff}$ in Equations (\ref{eq:peak}) and (\ref{eq:Aeff2}).

\subsection{Power-law Temperature Distributions}
\label{sec:multiT}

\begin{figure*}[!tbh]
\centering
\includegraphics[width=0.9\columnwidth]{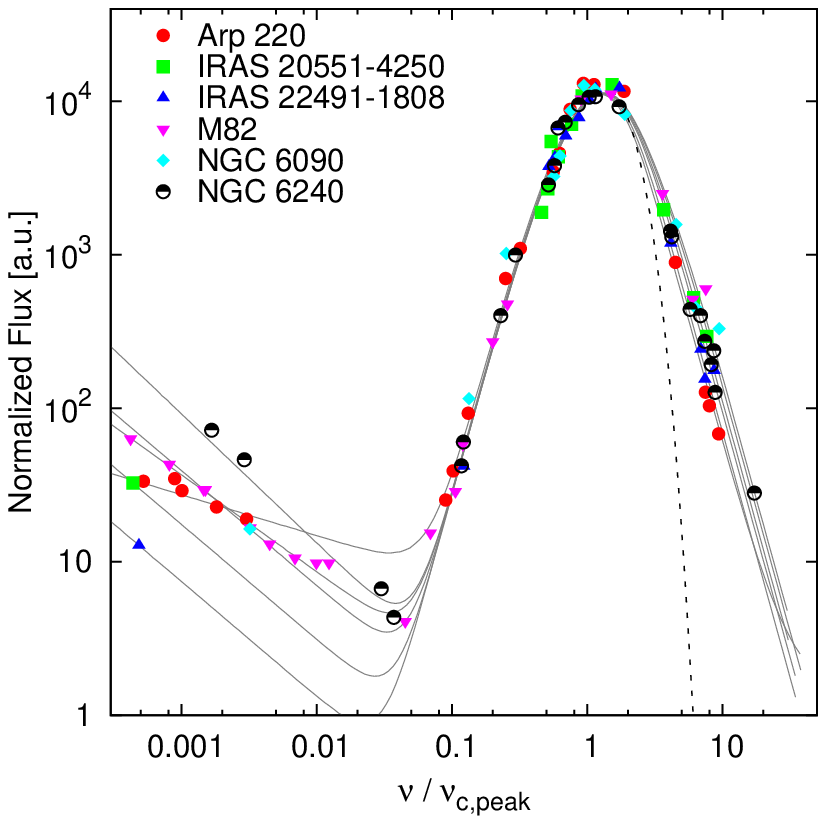}
\hskip 1cm
\includegraphics[width=0.9\columnwidth]{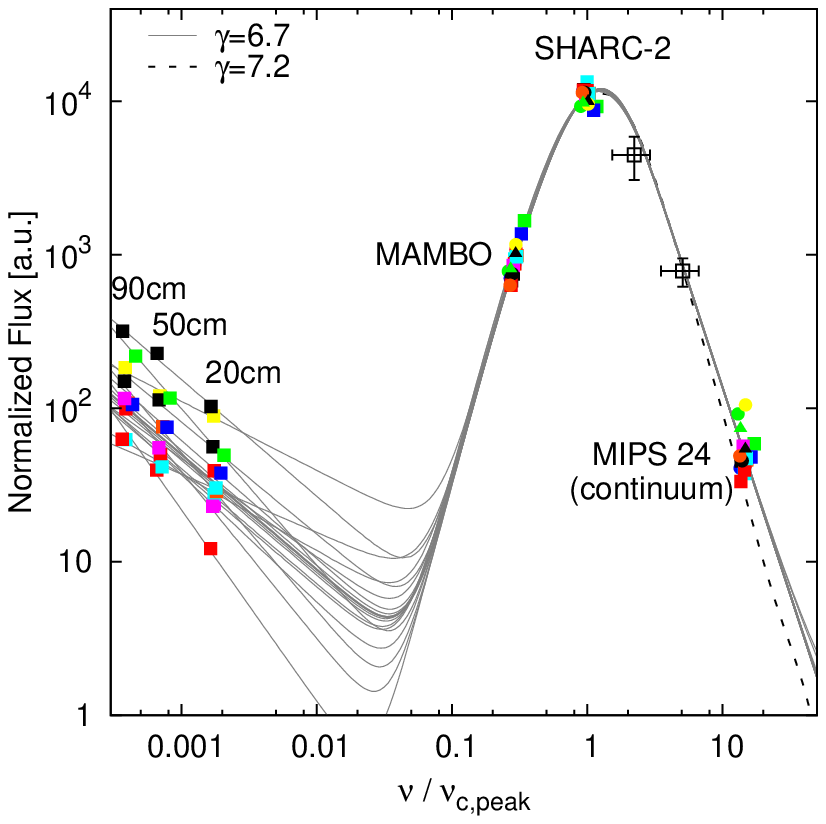}
\caption{Composite radio and IR continuum SEDs of local starbursts (left) and the SWIRE/MAMBO sample (right). The gray lines represent the best-fit gray-body models with power-law mass--temperature distributions using $\gamma$=7.2 for the local starbursts and $\gamma$=6.7 for the SWIRE-selected sample. Frequencies are normalized to the peak of the cold-component emission (Equation (\ref{eq:peak})). A single-$T$ model is shown in the left panel for comparison (dashed line), demonstrating the shortcoming of such models on the Wien side. The continuum contribution to the 24\um flux density is estimated based on the average IRS spectra of eight \citet{Lonsdale2009} galaxies. The average local starburst model with $\gamma$=7.2 is also shown in the right panel for comparison (dashed line). We also show the stacked average 70\um and 160\um flux densities of \citet{Fiolet2009} sources (black hollow squares with error bars) for a fictive source with median properties ($T_c$=33.5\,K, $M_d$=7$\times$10$^8$\,M$_\odot$ at $z$=2.05).The sources with insufficient radio data assume α = 0.75 for the radio spectral index. Individual SED plots for the local starbursts (part b) and the SWIRE-selected sample (part c) are available in the online edition. For part c, the 24 μm data points reflect the expected continuum contribution to the in-band flux-densities, according to Figure 3, with an uncertainty of 0.25 dex. (An extended color version of this figure is available in the online journal.) 
}
\label{fig:composite}
\end{figure*}

Single-temperature models can be useful, especially for offering comparison to earlier studies that analyzed data in this way. However, it is clear that such models fall short on the Wien side of the thermal emission spectrum (dotted line in left panel of Figure \ref{fig:composite}). Measurements at wavelengths shorter than the peak (typically $\simeq$200$\mathum$) by {\em IRAS}, {\em ISO}, or {\em Spitzer} all indicate that flux densities fall far less dramatically than single-$T$ gray-body models imply. Temperature estimates are biased by the inclusion of data on the Wien side (at $\lesssim$200\um in the rest frame), and luminosities are systematically underestimated. The errors can be significant depending on the actual spectral profile, and the data used for fitting. This is a concern, since the SHARC-2 350\um point translates to rest frame wavelengths of 90-175\um at redshifts z$\sim$1--3.

For improving our models, we look for clues in the same local starbursts that yield the oft-used templates for the interpretation of {\em Spitzer} data \citep{Berta2005, Polletta2007}. Based on these galaxies we find that a power-law distribution of temperature components ($d M_d/ d T \propto T^{-\gamma}$), with a low-temperature cutoff at $T_c$, are in excellent agreement with measurements (Figure \ref{fig:composite}). At the same time, power-law distributions are easily justified in a physical context. The radiative-transfer models of \citet{Dale2001} and \citet{DaleHelou2002} are similarly conditioned on dust being distributed in a range of radiation fields $U$ as $dM/dU \propto U^{-\alpha}$. Their power-law index $\alpha$ is also the Wien-side spectral index and depends on the geometric configuration of dust clouds relative to the heating sources. Under the assumption of radiative dust heating the two models are approximately equivalent with $\gamma \approx 4 + \alpha + \beta_{\rm eff}$. Therefore, we expect $\gamma$$\approx$6.5--7.5 for sources embedded in a diffuse medium (e.g., star formation), and $\gamma$$\approx$4--5 if absorption happens inside a dense medium at a well-defined distance from the source (e.g., for a molecular ring surrounding an AGN). 

We argue that a temperature distribution (as opposed to a radiation-field distribution) offers a more general treatment, because it can also account for non-radiative heating of dust particles, e.g., by shocks or gravitational collapse. One needs only to remember that one of the critical functions of dust in star formation is to act as a coolant during the collapse of star-forming gas clouds. Thus, dust heating begins even before the UV-bright OB stars illuminate the interstellar medium. Temperature distributions can further arise from a distribution of individual molecular clouds in a galaxy, each with its own size and environment.

Apart from the physical plausibility of power laws, a practical advantage is that they add just a single variable $\gamma$ to the parameterization of single-$T$ models. This is welcome, especially since the number of SED measurements across the FIR band are generally few, even for many well-studied local galaxies, allowing robust constraining only for a handful of parameters.

We can write appropriate expressions for predicted flux densities in terms of the single-$T$ flux $S_1$ (Equation (\ref{eq:S1})) as

\begin{equation}
S_{\nu_{\rm obs}}(T_c) = (\gamma-1)~T_c^{\gamma-1} \int_{T_c}^{\infty} S_1(\nu_{\rm obs}, T_{\rm obs})~T^{-\gamma}~dT, 
\end{equation}

where once again $T_{\rm obs} = T/(1+z)$. The integration can be performed numerically with a number of discrete temperature components, or can be written in a closed form \citep[see][]{Kolbig1970} using the incomplete Riemann zeta function $Z_1(\gamma-1, ~h \nu_r / k T_c)$ and $\Gamma(\gamma-1)$. Luminosities can be expressed in a closed form and are given by

\begin{equation}
L = \frac { \gamma - 1 } { \gamma - (5+\beta_{\rm eff}) } ~ L_1(T_c),
\end{equation}

based on the single-$T$ value $L_1$ (Equation (\ref{eq:L1})). Finiteness of luminosities requires that $\gamma$ be larger than 5+$\beta_{\rm eff}$. 

Our measure of the luminosity accounts for the full power radiated through thermal dust emission, in contrast to a more traditionally defined FIR or IR luminosities, which measure output inside a strictly defined wavelength ranges (e.g., between 40\um and 1000$\mathum$). As such, our expression for the integrated luminosity can be applied to a wider range of environments (e.g., hot accretion disks) than definitions based on strict wavelength ranges. Additionally, $L$ provides a fair proxy for the total bolometric luminosity of warm ($\gtrsim$30\,K) LIRGs, where an overwhelming fraction of the total power output is absorbed and re-radiated by dust.

\subsubsection{Clues from Local Starbursts}

To check what $\beta$ and $\gamma$ describe the local starbursts best, we fitted a selection\footnote{We relied on the online NASA Extragalactic Database (NED) for flux densities. When more than one measurement was available for a given band we generally opted for the most recent one, or the one with the lowest measurement uncertainty. Additionally, we supplemented the data for M\,82 in the 350$\mathum$--2\,mm range with \citet{Leeuw2009} and \citet{Alton1999}. For Arp\,220, we also included our own 350\um measurement of 9.83$\pm$0.11\,Jy in the analysis.} of published FIR continuum data uncontaminated by aromatic features or synchrotron emission, in the 6$\mathum$--2\,mm range, with power-law $T$ distributions and assuming kpc-scale emission. For M\,82 we have assumed a characteristic emission diameter of 300\,pc, in line with the observed size of its molecular ring. Our fitting routine performs a generic $\chi^2$ minimization based on the downhill simplex method described by \citet{Press1986}. The results are summarized in Table~\ref{tab:SB}.

Encouraged by the consistent $\gamma$ values within measurement error, we stipulate that the temperature distribution is possibly sufficiently universal for star-forming galaxies, so that we may proceed to fit $\gamma$ and $\beta$ globally, while keeping $T_c$ and $M_d$ as individual parameters for each galaxy. We do this by nesting an inner $\chi^2$ minimization loop for $T_c$ and $M_d$ for each objects, in an external fitting of $\beta$ and $\gamma$ for all. Thus we find that $\beta=1.53\pm0.04$ and $\gamma=7.22\pm0.09$ describe the local starburst best, producing close to the expected level of residual scatter based on the measurement and calibration uncertainties. The resulting $\beta$ is in excellent agreement with measurements both in the Galaxy \citep{Dupac2003} and in the laboratory \citep{Agladze1996}. Under the assumption of radiative dust heating, the values of $\gamma$ and the characteristic opacities $\tau$$\gtrsim$0.3 on the Wien side imply radiation-field index $\alpha$$\gtrsim$2.35, in good agreement with the predicted value of 2.5 for sources embedded in a diffuse medium \citep{Dale2001}. The conclusions on $\beta$ and $\gamma$ hold for a wide range of possible emission scales around the assumed values.

\subsection{Sample Characterization}

In characterizing our SWIRE-selected samples, we assume $\beta$=1.5 based on our findings on local starbursts and work by others. For the size of the emitting region (i.e., for setting $d\Omega_AD_A^2$), we assume an average diameter of 3\,kpc in our single-temperature fits. Such an extended starburst is unprecedented in the local universe, where the luminosities tend to originate from smaller kpc-scale molecular rings. Nevertheless, the larger sizes are reasonable for our sample \citep[see, e.g.][]{Farrah2008}, where luminosities exceed the local examples by 1 or 2 orders of magnitude. The assumption on the extent of emission is critical only for interpreting the temperature parameter of the fit as a true measure of physical temperatures. The best-fit temperature values would rise if smaller emission diameters were assumed. At the same time, dust masses and integrated luminosities remain nearly independent of the size assumption, and thus offer robust characterizations. The resulting single-temperature fits to the sample are summarized in Table~\ref{tab:singleT}.

The FIR properties derived from the single-$T$ fits are based on the MAMBO 1.2\,mm measurement and our SHARC-2 350\um data only. Shorter wavelength data (especially MIPS 24$\mathum$) were not included in these fits because these trace higher-temperature components and aromatic features, neither of which are accounted for in the single-$T$ dust model. We also do not include the 160\um datum of EN1-04 for any in our modeling, because we find it inconsistent with the measurements at 24\um and at 1.2\,mm. We suspect that the uniquely detected 160\um flux density of our sample may be contaminated by an unknown source close to EN1-04.

\begin{figure}[!b]
\centering
\includegraphics[width=\columnwidth]{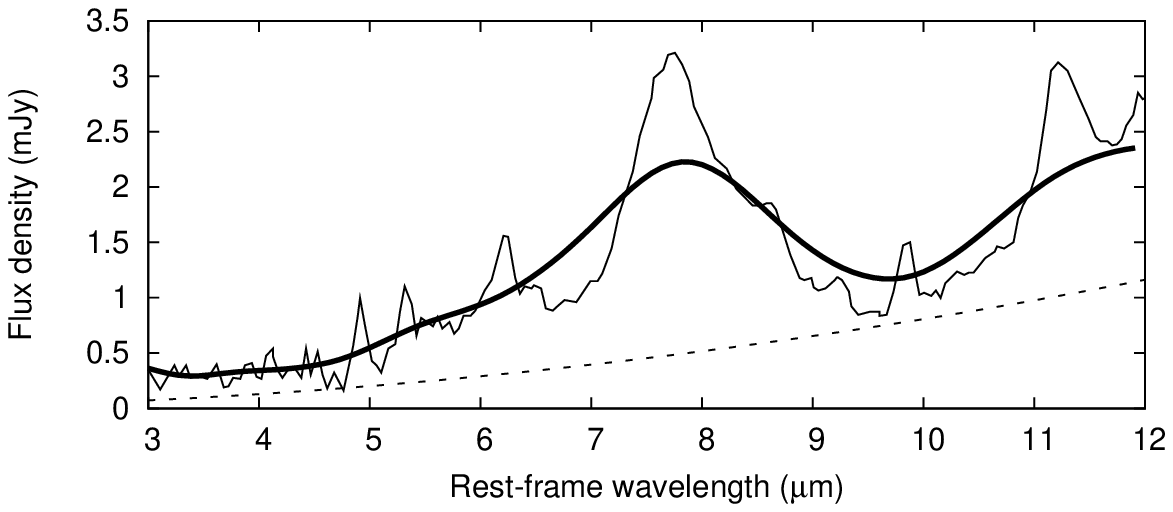}
\hfill
\includegraphics[width=\columnwidth]{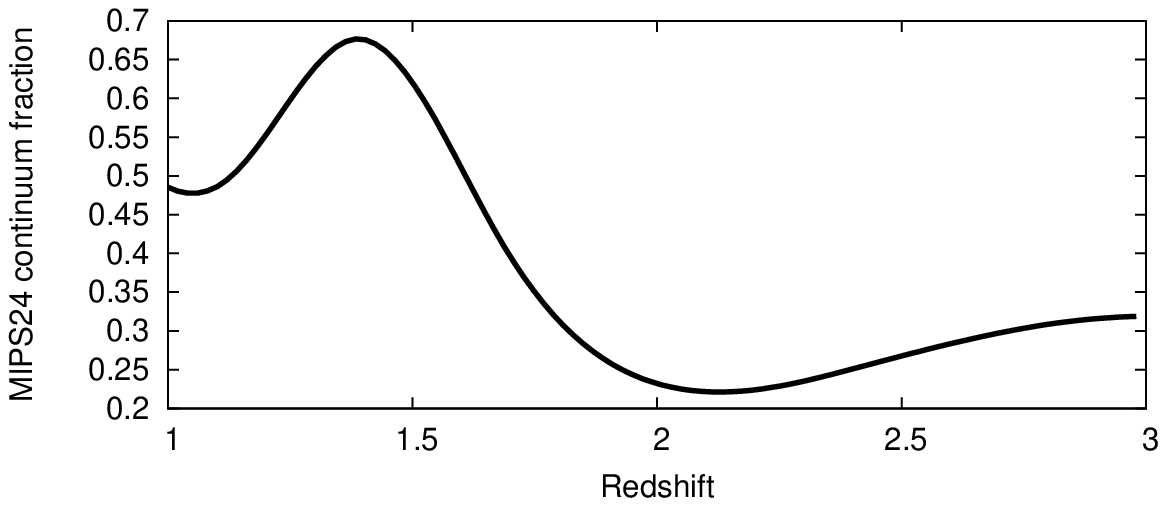}
\caption{Expected continuum contribution to the MIPS 24\um measurements. The top panel shows the average IRS spectra of eight \citet{Lonsdale2009} sources (thin solid line), and as it would appear under the resolution of the MIPS 24\um band (thick solid curve). Based on the decomposition of this spectrum into starburst PAH features (NGC3079) and a power-law continuum (dashed line) contribution $\sim$$\lambda^2$ \citep[see][]{Lonsdale2009}, we estimate the typical continuum contributions to the MIPS 24\um flux as a function of redshift (bottom panel) for the SWIRE-selected samples of \citet{Lonsdale2009} and \citet{Fiolet2009}, which we use to derive appropriate continuum constraints at 24\um for our SED fits.}
\label{fig:contrib24}
\end{figure}

Even with all the caution we can summon, single-temperature models cannot yield accurate characterizations for reasons already disserted. We include these simple models only to provide a comparison to other studies \citep[especially][]{Kovacs2006}, which analyze FIR and (sub)millimeter data in this way. Fitting a distribution of dust temperatures to the data yields more realistic estimates of the dust masses and luminosities. Our best estimates of the underlying FIR properties are summarized in Table~\ref{tab:multiT}, and shown in Figure~\ref{fig:composite}. The temperature parameter $T_c$ indicates the temperatures of the dominant coldest component, above which warmer components exist in lesser quantities (see Section~\ref{sec:multiT}). Therefore, it is not surprising that single-component fits, which measure an average temperature, should yield characteristic values above the cold cutoff temperatures $T_c$ in a distribution. 

For our temperature-distribution fits, we have included the MIPS 24\um data as guiding constraints on the short-wavelength tails of the dust continuum. It is clear, however, that the 24\um flux density is likely dominated by emission from aromatic features. However, warm dust and PAH likely cohabit the same interstellar environments. Thus, PAH emission is almost certainly coupled to a comparably bright dust continuum. Consequently, one should anticipate close to constant PAH-to-continuum flux ratios for a class of star formation dominated objects. Based on the stacked IRS spectra of \citet{Lonsdale2009} we can estimate the typical continuum contribution to the MIPS 24\um measurement as a function of redshift (Figure~\ref{fig:contrib24}), with which we set appropriate 24\um constraints for our fits. Recognizing, however that the continuum fraction may vary significantly from one galaxy to the next, we assign an uncertainty of 0.25 dex to these 24\um guesstimates. Thus, we allow the 24\um continuum fraction to span a full decade in brightness before the deviation becomes significant (i.e., $>$2\,$\sigma$), while we assume that the average IRS spectrum of \citet{Lonsdale2009} is representative of the sample as a whole.

\section{Discussion}
\label{sec:discussion}

Under the assumption of extended starbursts we find that our models deviate from the optically thin limit only moderately, with $\tau$$\approx$0.7$\pm$0.2 near the emission peaks (under the assumed emission diameters of 3\,kpc).  This is good news for a straightforward comparison to the SMGs of \citet{Kovacs2006}, which were analyzed under the optically thin assumption. These optical depths are not much higher than the opacities of local starbursts on kpc scales. We may take this as a first indication that our guess on the extent of star formation is correct. 

The median 70\um and 160\um flux densities of 2.56$\pm$0.38\,mJy and 14.63$\pm$4.05\,mJy, respectively, from the stacked images of the MAMBO-detected sources of \citet{Fiolet2009} agree well with the multi-$T$ models of our sources with median properties (i.e., $T_c$=33.5\,K, $M_d$=10$^{8.82}$\,M$_{\odot}$ at $z$=2.05), within the expected measurement uncertainties (see Figure~\ref{fig:composite}). These prove the general reliability of the 24\um continuum contribution estimates.

The use of the scaled 24\um data further allows us to probe the range of $\gamma$ values and typical emission scales that are consistent with measurements. Accordingly, we find $\gamma$=6.71$\pm$0.11 and average emission diameters around $\simeq$2\,kpc fit the data best globally. We can exclude diameters $<$1.2\,kpc with 95\% confidence. Thus, we see our size assumption for the single-$T$ fits sufficiently justified. At the same time, we adopt the most likely diameter of 2\,kpc in our multi-$T$ fits.

The result on the sizes is significant because it provides an indirect measure of the true extent of the FIR emission in hyper-luminous infrared galaxies (HLIRGs). Interferometric imaging by \citet{Tacconi2006} already hinted at similar scales based on slightly broadened CO emission profiles (versus their synthesized beams). The broadening however can also result from distinct compact nuclei separated at the inferred distance (e.g., in a merging system), whereas the best-fit SEDs are only consistent with extended dust emission. The implied scales, spanning kiloparsecs across, are in line with \citet{Farrah2008} and are a strong indication that dust heating is distributed, which in turn supports star formation as the primary power source. 

\begin{deluxetable}{@{\extracolsep{0pt}} l \pmcol \pmcol \pmcol \pmcol}[!tb]
\tablewidth{\columnwidth}
\tablecolumns{9}
\tablecaption{Single-temperature Characterizations\label{tab:singleT}}
\tablehead{
\colhead{}   & \cc{$T$} & \cc{$\log M_d$} & \cc{$\log L_1$}  & \cc{} \\ 
  ID & \cc{(K)} & \cc{($M_\odot$)}  & \cc{($L_\odot$)} & \cc{$q$} 
}
\startdata
LH-02  & 36.5&2.7 & 9.13&0.15 & 12.66&0.10 & \nopmdata \\
LH-06  & 33.7&3.7 & 9.03&0.19 & 12.46&0.15 & \nopmdata \\
LH-03  & 53.5&7.7 & 8.66&0.20 & 13.23&0.21 & \nopmdata \\
L-1    & 39.1&4.9 & 8.79&0.20 & 12.64&0.17 & 2.16&0.18 \\
L-9    & 41.3&6.4 & 8.90&0.17 & 12.82&0.24 & 2.41&0.25 \\
L-11   & 37.5&4.3 & 8.85&0.18 & 12.59&0.15 & 1.83&0.16 \\
L-14   & 40.4&7.3 & 8.63&0.28 & 12.61&0.22 & 1.94&0.22 \\
L-15   & 33.9&4.3 & 8.81&0.22 & 12.34&0.15 & 1.99&0.16 \\
L-17   & 43.3&8.9 & 8.61&0.26 & 12.75&0.29 & 1.76&0.29 \\
L-20   & 46.9&8.8 & 8.63&0.26 & 12.93&0.25 & 2.48&0.26 \\
L-21   & 36.0&5.1 & 8.89&0.22 & 12.52&0.19 & 2.17&0.20 \\
L-22   & 33.4&3.5 & 8.97&0.18 & 12.41&0.15 & 1.56&0.16 \\
L-23   & 32.8&5.7 & 8.95&0.24 & 12.36&0.26 & 1.80&0.27 \\
L-25   & 36.3&5.3 & 8.79&0.22 & 12.48&0.18 & 2.07&0.19 \\
L-27   & 36.3&5.6 & 8.77&0.25 & 12.47&0.20 & 2.30&0.21 \\
LH-01  & 27.1&2.0 & 9.35&0.13 & 12.16&0.11 & \nopmdata \\
EN1-01 & 45.6&7.7 & 8.66&0.24 & 12.89&0.23 & \nopmdata \\
EN1-02 & 43.2&9.4 & 8.65&0.29 & 12.77&0.30 & \nopmdata \\
EN1-04 & 33.6&6.2 & 8.89&0.28 & 12.38&0.26 & \nopmdata \\
EN2-01 & 34.5&4.9 & 8.82&0.23 & 12.39&0.18 & \nopmdata
\enddata
\tablecomments{All quantities were fitted using a single-temperature model with $\beta=1.5$ and assuming a 3\,kpc emission diameter. The dust masses assume $\kappa(850\mathum)$=0.15\,m$^2$\,kg$^{-1}$. Uncertainties are 1$\sigma$ total errors of the fits to data, which do not include the uncertainties in the redshift values.}
\end{deluxetable}

For star-forming galaxies, we expect $\gamma$$\approx$6.5+$\beta_{\rm eff}$ based on \citep{Dale2001}. With $\tau_{\rm pk}$$\gtrsim$1.2$\pm$0.4 near the emission peaks (corresponding to the most likely emission diameters of 2\,kpc), it is not surprising that the SWIRE-selected sample has $\gamma$ values below those of local starbursts. The relatively shallow temperature distributions in turn might explain the unusually high 24\um to 1.2\,mm flux density ratios reported by \citet{Lonsdale2009} and \citet{Fiolet2009}. The implied radiation-field distribution indices $\alpha$$\gtrsim$2.0--2.35 further support star formation as the primary heating source behind the extreme luminosities.

The average dust mass in our sample is 7$\times$10$^8$\,M$_\odot$. All of our galaxies have masses near the mean value, within the typical 40\% (i.e., 0.15\,dex) measurement error, revealing only a hint of an underlying scatter $\lesssim$20\%. Assuming SMG-like gas-to-dust mass ratios of 54 \citep{Kovacs2006}, which are also typical of the nuclear regions of nearby galaxies \citep{Seaquist2004}, we estimate an average molecular gas content for our sample in the neighborhood of 4$\times$10$^{10}$\,M$_\odot$.

Our analysis reveals no statistical difference between the \citet{Lonsdale2009} and \citet{Fiolet2009} subsamples. Table~\ref{tab:comparison} offers a comparison of average properties between the subsamples, the combined SWIRE/MAMBO sources, and the classical SMGs.
The results from the single-$T$ fits are statistically indistinguishable from the values found for classical SMGs by \citet{Kovacs2006}. 
As such, we tentatively conclude that their samples consist predominantly of classical SMGs, although with a more restricted range of redshifts ($z$$\simeq$2.05$\pm$0.30) than the purely submillimeter-selected samples. The narrow redshift distribution is a direct product of our selection criteria, which favor objects close to $z$$\approx$2. Since the radio co-selected SMG population peaks around $z$$\approx$2.3 \citep{Chapman2003,Chapman2005}, we conclude that two-pronged SWIRE selection (seeking the 24$\mathum$ bright population of 5.8\um peakers) is effective in identifying a representative subset of SMGs at mid-IR wavelengths.
 

\begin{deluxetable}{@{\extracolsep{0pt}} l \pmcol \pmcol \pmcol \pmcol }[!tb]
\tablewidth{\columnwidth}
\tablecolumns{9}
\tablecaption{Realistic Characterizations\label{tab:multiT}}
\tablehead{
\colhead{}   & \cc{$T_c$} & \cc{$\log M_d$} & \cc{$\log L$}  & \cc{} \\ 
  ID & \cc{(K)} & \cc{($M_\odot$)}  & \cc{($L_\odot$)} & \cc{$q$} 
}
\startdata
LH-02  & 34.6&1.9 & 9.07&0.13 & 12.97&0.08 & \nopmdata \\
LH-06  & 33.5&2.9 & 8.86&0.18 & 12.93&0.12 & \nopmdata \\
LH-03  & 45.0&3.4 & 8.71&0.12 & 13.44&0.13 & \nopmdata \\
L-1    & 35.6&2.8 & 8.73&0.16 & 13.04&0.12 & 2.25&0.15 \\
L-9    & 34.8&3.0 & 8.93&0.12 & 12.99&0.15 & 2.28&0.16 \\
L-11   & 33.2&2.4 & 8.84&0.13 & 12.91&0.11 & 1.85&0.13 \\
L-14   & 33.5&2.7 & 8.69&0.16 & 12.93&0.12 & 1.97&0.15 \\
L-15   & 29.8&2.3 & 8.78&0.16 & 12.73&0.11 & 2.08&0.14 \\
L-17   & 34.1&2.9 & 8.71&0.14 & 12.96&0.14 & 1.67&0.16 \\
L-20   & 36.7&2.7 & 8.77&0.13 & 13.09&0.11 & 2.34&0.14 \\
L-21   & 31.1&2.4 & 8.91&0.15 & 12.80&0.12 & 2.15&0.15 \\
L-22   & 33.1&2.7 & 8.79&0.21 & 12.91&0.11 & 1.76&0.15 \\
L-23   & 33.0&2.7 & 8.78&0.21 & 12.90&0.14 & 2.04&0.17 \\
L-25   & 32.7&2.7 & 8.74&0.17 & 12.89&0.13 & 2.17&0.16 \\
L-27   & 31.9&2.7 & 8.75&0.17 & 12.84&0.13 & 2.38&0.16 \\
LH-01  & 27.7&1.7 & 9.21&0.15 & 12.58&0.09 & \nopmdata \\
EN1-01 & 38.5&3.1 & 8.70&0.15 & 13.17&0.13 & \nopmdata \\
EN1-02 & 34.7&3.0 & 8.75&0.15 & 12.99&0.14 & \nopmdata \\
EN1-04 & 32.0&3.0 & 8.77&0.22 & 12.85&0.15 & \nopmdata \\
EN2-01 & 32.5&2.9 & 8.70&0.18 & 12.88&0.13 & \nopmdata
\enddata
\tablecomments{Same as Table~\ref{tab:singleT}, except that properties are derived for a power-law temperature distribution with $\gamma=6.7$, and a most-likely value of 2\,kpc is assumed for the emission diameter. We used $\kappa(850\mathum)$=0.15\,m$^2$\,kg$^{-1}$ and $\beta$=1.5 in deriving the quantities.}
\end{deluxetable}

\begin{deluxetable*}{@{\extracolsep{0pt}} l \pmcol \pmcol \pmcol \pmcol \pmcol \pmcol \pmcol \pmcol \pmcol}[!tbh]
\tablewidth{\textwidth}
\tablecolumns{19}
\tablecaption{Comparison to Classical SMGs\label{tab:comparison}}
\tablehead{
  \colhead{} & \cc{$T_1$} &  \cc{$T_c$} & \cc{$\log M_1$} & \cc{$\log M_d$} & \cc{$\log L_1$}  & \cc{$\log L$} & \cc{}  & \cc{} & \cc{} \\ 
  Sample & \cc{(K)} &  \cc{(K)} & \cc{($M_\odot$)}  & \cc{($M_\odot$)} & \cc{($L_\odot$)} & \cc{($L_\odot$)} & \cc{$q$} & \cc{$\tau_{\rm pk}$} & \cc{$\gamma$} 
}
\startdata
local starbursts\tablenotemark{$a$} & 
\nopmdata & 34.7&3.6  & \nopmdata & 7.33&0.62 & \nopmdata  & 11.82&0.51 & 2.42&0.35 & 0.3&0.2 & 7.22& 0.09\\
\citet{Lonsdale2009} & 
36.1&5.8  & 34.4&4.2  & 8.96&0.18 & 8.85&0.12 & 12.59&0.28 & 12.97&0.22 & \nopmdata & 1.4&0.5 & \nopmdata \\
\citet{Fiolet2009} & 
\cc{36.5} & \cc{33.1} & \cc{8.83} & \cc{8.80} & \cc{12.54} & \cc{12.91} & 2.08&0.16 & 1.1&0.2 & \nopmdata \\
\citet{Lonsdale2009} $+$ \citet{Fiolet2009} &
35.5&2.2  & 33.5&2.1  & 8.90&0.07 & \cc{8.82} & 12.56&0.05 & 12.93&0.12 & \nopmdata & 1.2&0.4 & 6.71&0.11 \\
SMGs        &
34.1&9.5  & (30.4&8.5)\tablenotemark{$b$} & 8.96&0.32 & (8.85&0.18)\tablenotemark{$b$} & 12.60&0.48 & (12.76&0.61)\tablenotemark{$b$} & 2.12&0.12 & (1.2&0.5)\tablenotemark{$b$}  & \nopmdata \\
SMGs (1.5$<$$z$$<$2.5) &
32.1&5.6  & (30.8&5.5)\tablenotemark{$b$} & 9.07&0.26 & (8.87&0.11)\tablenotemark{$b$} & 12.57&0.30 & (12.82&0.35)\tablenotemark{$b$} & 2.05&0.21 & (1.2&0.4)\tablenotemark{$b$}  & \nopmdata
\enddata
\tablecomments{A comparison of our {\em Spitzer}-selected samples to the local starbursts, and the  classical SMGs of \citet{Kovacs2006}.
Indicated are the ensemble average values and 1\,$\sigma$ intrinsic scatters for the characteristic dust temperature (i.e., single-$T$ temperature) $T_1$, cold-component temperature $T_c$, single-$T$ dust mass $M_1$, dust mass $M_d$, single-$T$ luminosity $L_1$, total (IR) luminosity $L$, the (F)IR--radio correlation measure $q$, and the characteristic opacity $\tau_{\rm pk}$ near the peak of the emission. Only mean values are shown, when the intrinsic scatters are not detected. The last column shows the average mass--temperature distribution index $\gamma$ and its rms uncertainty. The dust masses assume $\kappa(850\mathum)$=0.15\,m$^2$\,kg$^{-1}$.}
\tablenotetext{$a$}{These are the local starbursts of Table~\ref{tab:SB}, which provide the SWIRE starburst templates \citep{Berta2005, Polletta2007}.}
\tablenotetext{$b$}{Based on our own multi-$T$ fits to the \citet{Kovacs2006} data and assuming $\gamma$=6.7.}
\end{deluxetable*}

\subsection{Radio-FIR Correlation}
\label{sec:correlation}

Our dust fits point to a tight correlation between IR and radio luminosities. This is witnessed by the consistent $q$ values in Tables~\ref{tab:singleT} and \ref{tab:multiT}, which were calculated according to Equation (\ref{eq:q}).  Such correlations are well known for various local galaxies \citep[e.g.][]{Helou1988, CondonBroderick1991, Condon1992, Yun2001}, and extend to various populations into the distant universe \citep[see][]{Garrett2002, Appleton2004, Beelen2006, Kovacs2006}. A flux density-based measure of the correlation was originally formulated by \citet{Helou1985} using 1.4\,GHz radio and {\em IRAS} 60\um and 100\um flux densities. However, since we do not have the data in the rest-frame {\em IRAS} bands to use with the original definition, we can either guess these from our models, or redefine the correlation in terms of the quantities we can measure more directly.

\begin{figure}[!b]
\centering
\includegraphics[width=\columnwidth]{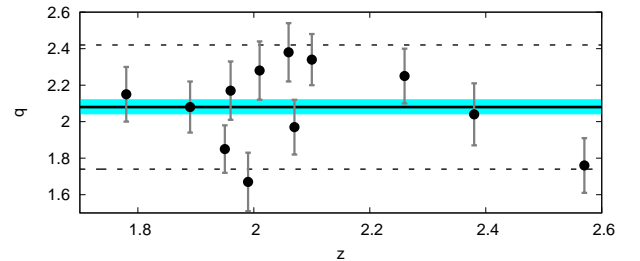}
\caption{Radio to (far-)infrared correlation. We plot the $q$ values for our sample, computed according to Equation (\ref{eq:q}) using the luminosities from Table~\ref{tab:multiT} and $\nu_n$=9\,THz. Also shown are the best-fit ensemble average value of 2.08$\pm$0.04 (solid line with shaded band), and the estimated 2$\sigma$ underlying scatter of data around it (dotted lines).
} 
\label{fig:q}
\end{figure}

The existence of the correlation is not surprising as both the synchrotron emission and the IR luminosities are driven by the short-lived population of high-mass ($>$8\,M$_\odot$) stars. The synchrotron emission arises when relativistic electrons, which are produced by supernovae, decelerate in the magnetic fields of galaxies. At the same time, the IR heating is dominated by the luminosities of massive OB stars. With short-lived massive stars responsible for both, measures based on luminosities, not flux densities, are expected to capture the essence of the correlation more accurately. Unlike many prior studies of the correlation, we have the advantage of well-constrained, precise luminosity estimates. Accordingly, we can express the logarithmic correlation constant as

\begin{equation}
q = \log \left( \frac {L } { \nu_n ~\mathcal{L}_{1.4\,{\rm GHz}} } \right),
\label{eq:q}
\end{equation}

with an appropriately chosen normalization frequency $\nu_n$ for the IR luminosity. The rest-frame radio luminosity density $\mathcal{L}_{1.4\,{\rm GHz}}$ is calculated from the observed flux density $S_{1.4\,{\rm GHz}}$ using:

\begin{equation}
\mathcal{L}_{1.4\,{\rm GHz}} = 4 \pi D_L^2~ (1+z)^{1-\alpha}~ S_{1.4\,{\rm GHz}}
\end{equation}

in terms of the luminosity distance $D_L$ and radio spectral index $\alpha$. The analysis of the SWIRE/MAMBO sample uses $\alpha$ values quoted in \citet{Fiolet2009}, while for the local starbursts we fitted $\alpha$ along with the other parameters.

It is practical to choose the normalization frequency $\nu_n$ such that the resulting $q$ values can be compared directly to the widely used original measure of \citet{Helou1985}, which assumed typical temperatures of 40\,K. We pick $\nu_n$ such that the definitions match exactly at that temperature, giving $\nu_n$$\rightarrow$4.52\,THz used with $L_1$ \citep[see][]{Kovacs2006}, or $\nu_n$$\rightarrow$9.0\,THz used with $L$ assuming $\gamma$=7.2. Consequently, our $q$ values closely mimic the \citet{Helou1985} definition. The correlation measures $q$ for our single- and multi-temperature fits are listed in Tables~\ref{tab:singleT} and \ref{tab:multiT}, respectively.

A different definition was suggested recently by \citet{Ivison2009}, who normalize IR luminosities, measured in the 8--1000\um band, with $\nu_n$=3.75\,THz intended for the 40--120\um band by \citet{Helou1985}. Their measure no longer offers a direct comparison to many earlier studies, which is why we prefer ours. However, our $q$ values can be compared to \citet{Ivison2009} after adding 0.38 (i.e., $\log [9.0/3.75]$).


We find best-fit average $q$ values of 2.11$\pm$0.08 with the single-temperature dust models and 2.08$\pm$0.04 for the power-law temperature distributions. The implied intrinsic scatters, over the measurement uncertainties, are estimated at 0.18 and 0.16, respectively. Both our mean values and intrinsic scatters (Figure~\ref{fig:q}) are significantly below those in local studies, and comparable to the SMGs analyzed by \citet{Kovacs2006}, and to the $q_{\rm IR}$$\sim$2.40 found by \citet{Ivison2009} for their correlation measure. The result further supports the conclusion that our sample primarily consists of classical SMGs, without significant excess heating either by an AGN, or by low-mass stellar populations (which are likely partially responsible for the higher average $q$ values observed in the local universe).

\section{Conclusions}

Our main conclusion is that the mid-IR selection criteria of the \citet{Lonsdale2009} and \citet{Fiolet2009}, picking the bright 24\um sources that are also 5.8\um peakers, is effective in identifying a significant fraction of the SMGs around $z$$\simeq$2, thus providing a means of enhancing our understanding of the elusive SMG population through studies at mid-IR and shorter wavelengths. The FIR characteristics (dust masses, temperature, and IR luminosities) of the SWIRE-selected sources are essentially identical to those of the classical SMGs at the same redshift. Additionally, the SWIRE-selected sample exhibits a correlation between radio and IR luminosities that is indistinguishable from that seen for SMGs. Our other conclusions are as follows.

\begin{enumerate}

\item{To provide realistic FIR characterizations for the sample, we developed new models for the SEDs of galaxies with power-law mass distributions of temperature components. We demonstrated that such models describe local starbursts extremely well, under a tightly constrained dust emissivity index $\beta$$\simeq$1.5 and mass-temperature index $\gamma$$\simeq$7.2. The $\gamma$ values are consistent with what we expect for heating dominated by star formation.}

\item{We find that the most likely IR spectral continuum shapes of the distant SWIRE-selected starbursts differ only slightly from the local examples, with best-fit $\gamma$$\simeq$6.7$\pm$0.1. The lower $\gamma$ values are consistent with star formation under the high optical depths ($\tau_{\rm pk}$$\simeq$1.2) in our sample.}
\item{Relying on the stacked IRS spectra of \citet{Lonsdale2009} for estimating the 24\um continuum contribution, we place the typical diameter of starburst activity in these SMGs above 1.2\,kpc (with 95\% confidence), and find a most likely value of 2\,kpc.}
\item{The emission scales, the typical $\gamma$, and the consistently low $q$ values all confirm that star formation is the main power source in the galaxies of the sample.}
\item{One of our targets shows a close clustering of three or four 350\um components, each of which contributes comparably to the observable 350\um flux density under the 15''--30'' resolution of most current (sub)millimeter surveys. A commonality of unresolved SMG multiplets could bias our understanding of this population. We need further studies to explore the importance of close multiplets in current and future submillimeter surveys.}

\end{enumerate}

\acknowledgements
The authors are extremely grateful to Tom Phillips for contributing extra observing time when we needed it most. The CSO is funded by the NSF Cooperative Agreement AST-083826. The Dark Cosmology Centre is funded by the DNRF. TRG acknowledges support from IDA.\\

{\it Facilities:} \facility{CSO (SHARC-2)}

\pagebreak

\clearpage
\pagebreak

\renewcommand{\thefigure}{\arabic{figure}\alph{subfigure}}
\setcounter{subfigure}{1}

\addtocounter{figure}{-3}
\addtocounter{subfigure}{1}

\begin{figure*}
\centering
\includegraphics[width=0.32\textwidth]{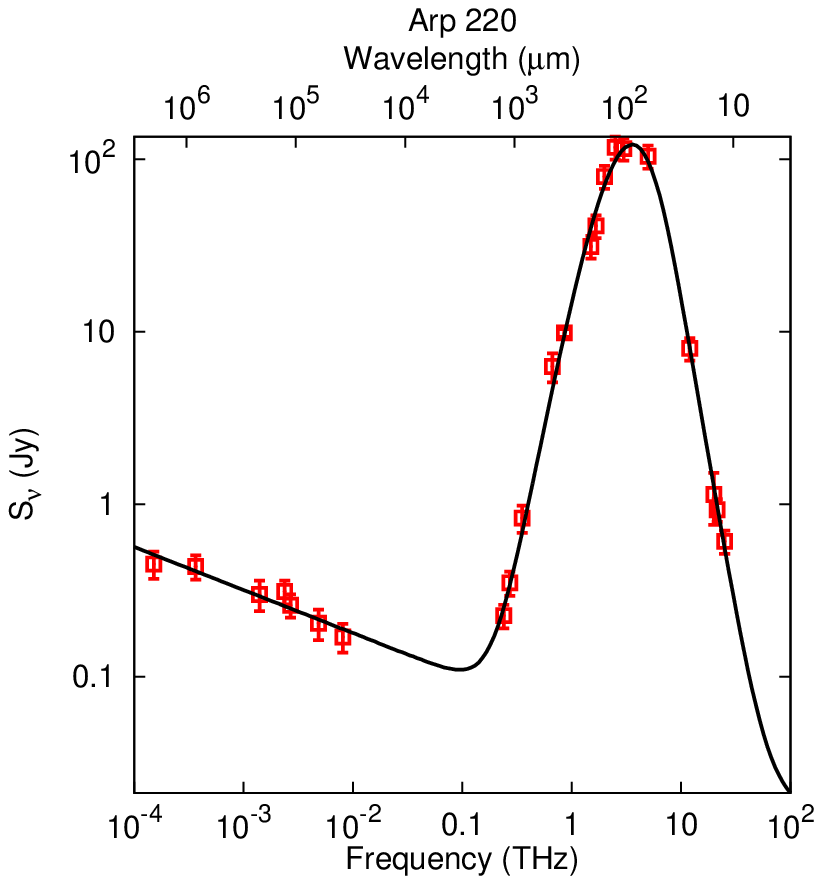}
\includegraphics[width=0.32\textwidth]{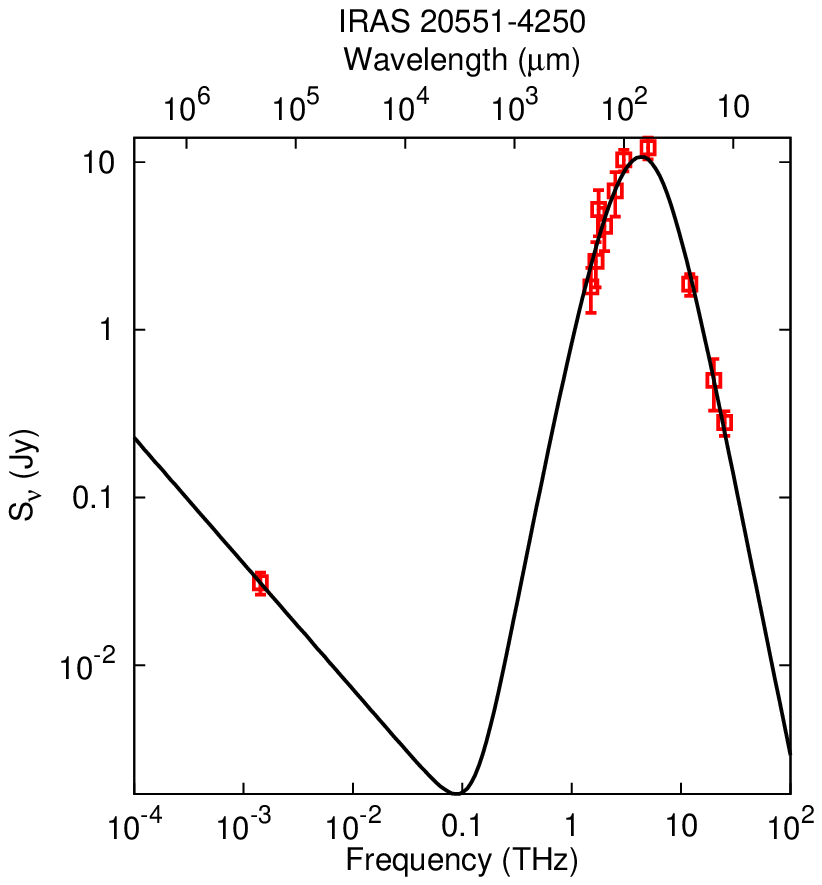}
\includegraphics[width=0.32\textwidth]{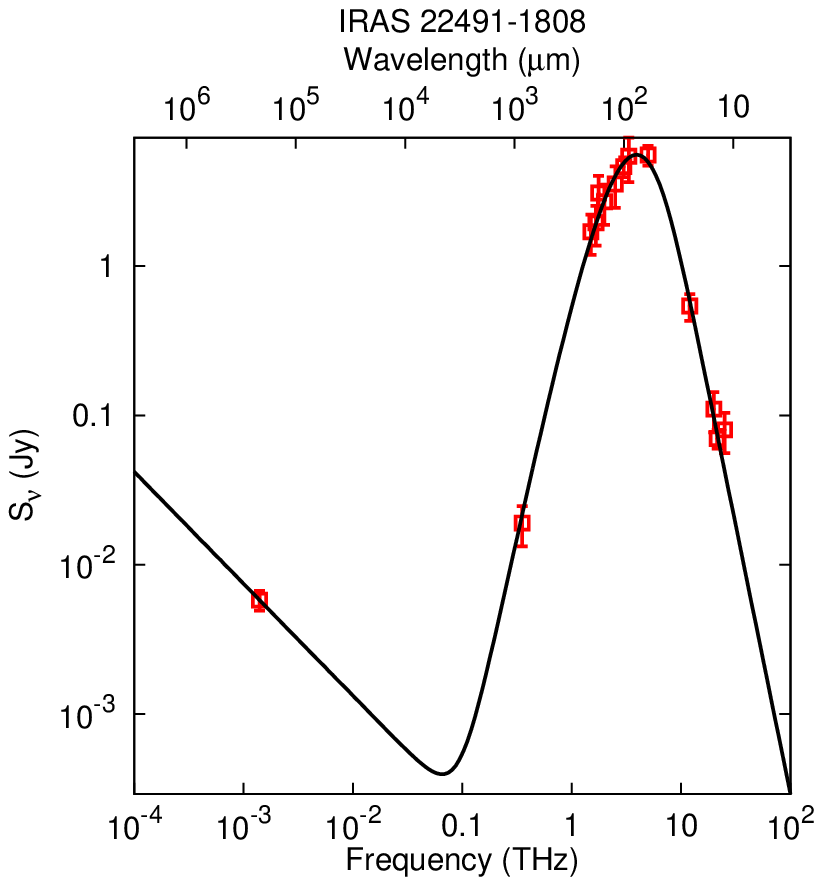}\\
\vspace{6pt}
\includegraphics[width=0.32\textwidth]{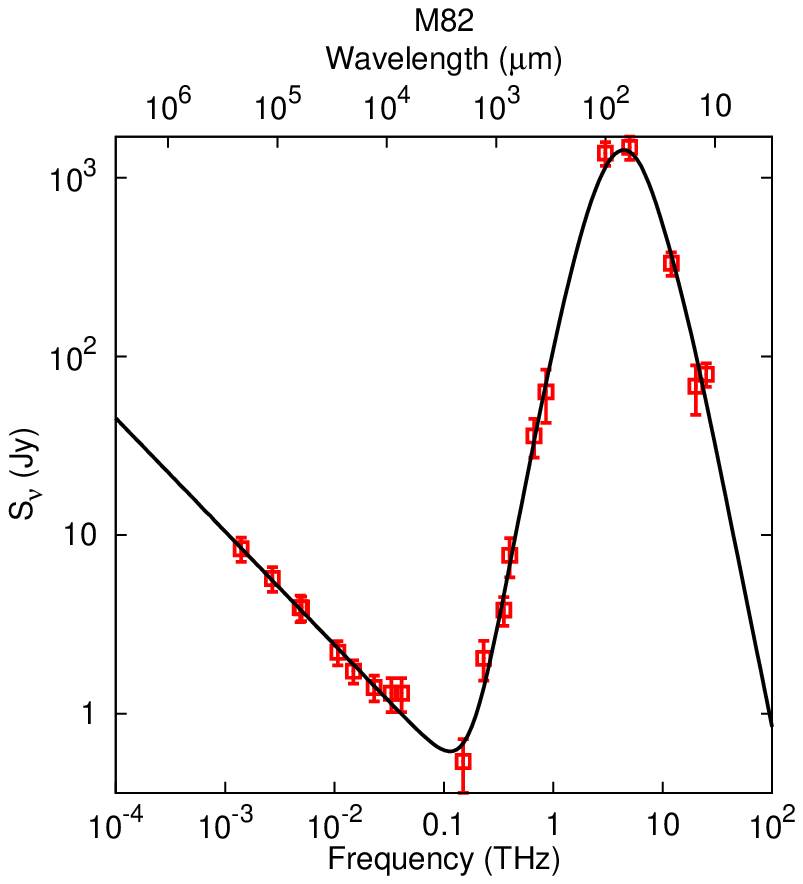}
\includegraphics[width=0.32\textwidth]{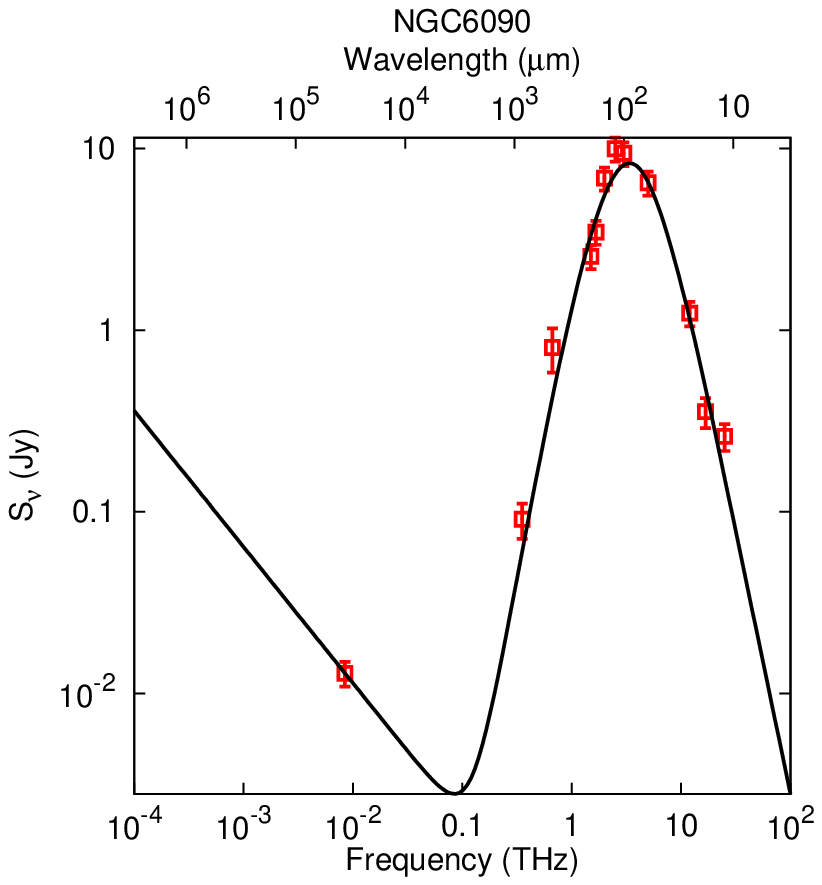}
\includegraphics[width=0.32\textwidth]{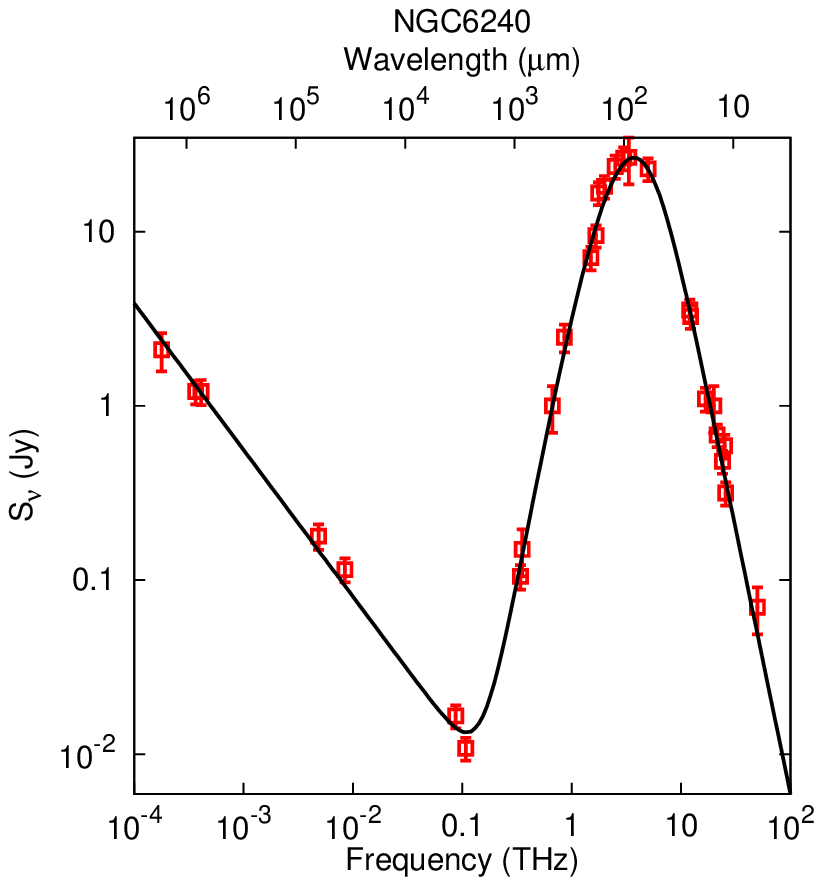}\\
\vspace{6pt}
\caption{Individual power-law temperature distribution SED fits to the local starbursts. The data used in our analysis are shown with hollow red squares and error bars. The sources with insufficient radio data assume $\alpha$=0.75 for the radio spectral index.
} 
\label{fig:localsed}
\end{figure*}

\addtocounter{figure}{-1}
\addtocounter{subfigure}{1}

\begin{figure*}
\centering
\includegraphics[width=0.23\textwidth]{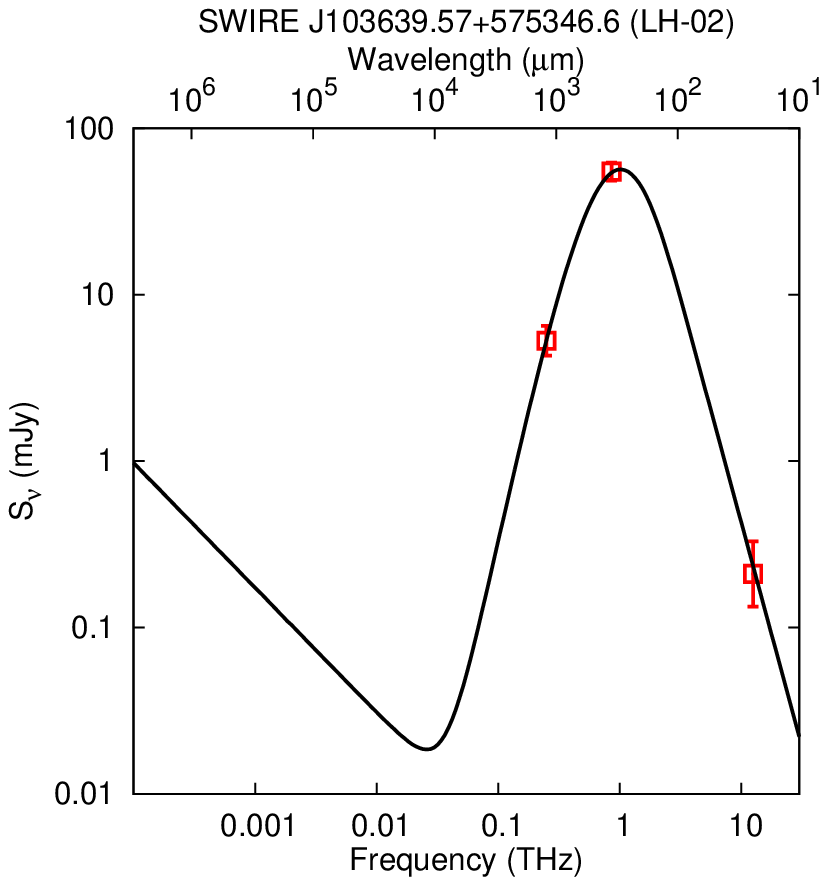}
\includegraphics[width=0.23\textwidth]{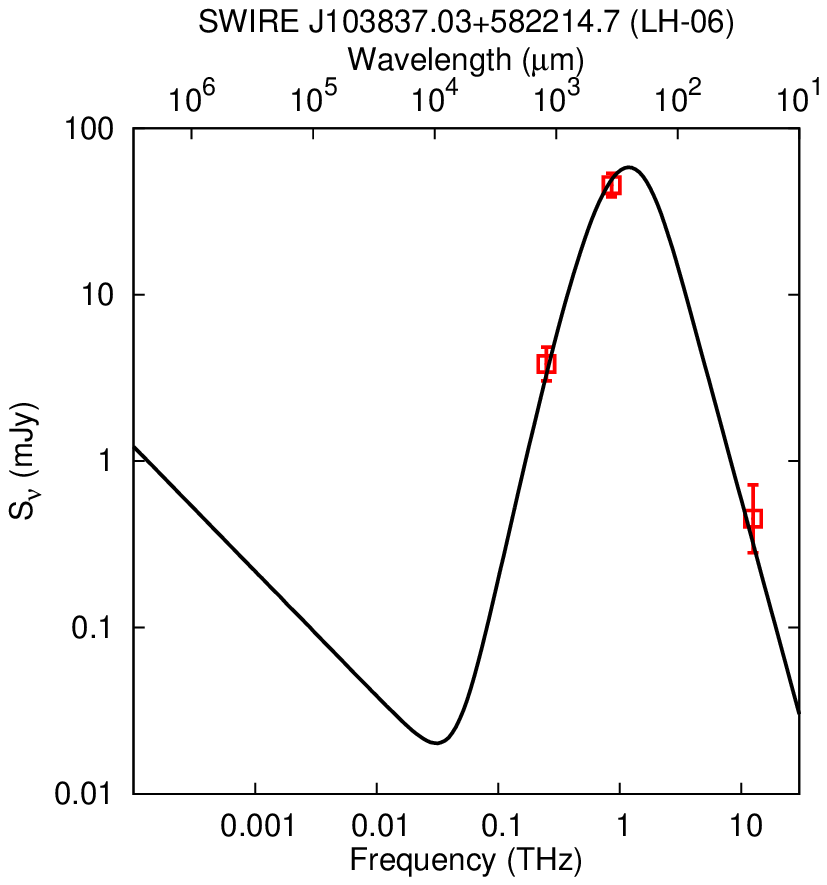}
\includegraphics[width=0.23\textwidth]{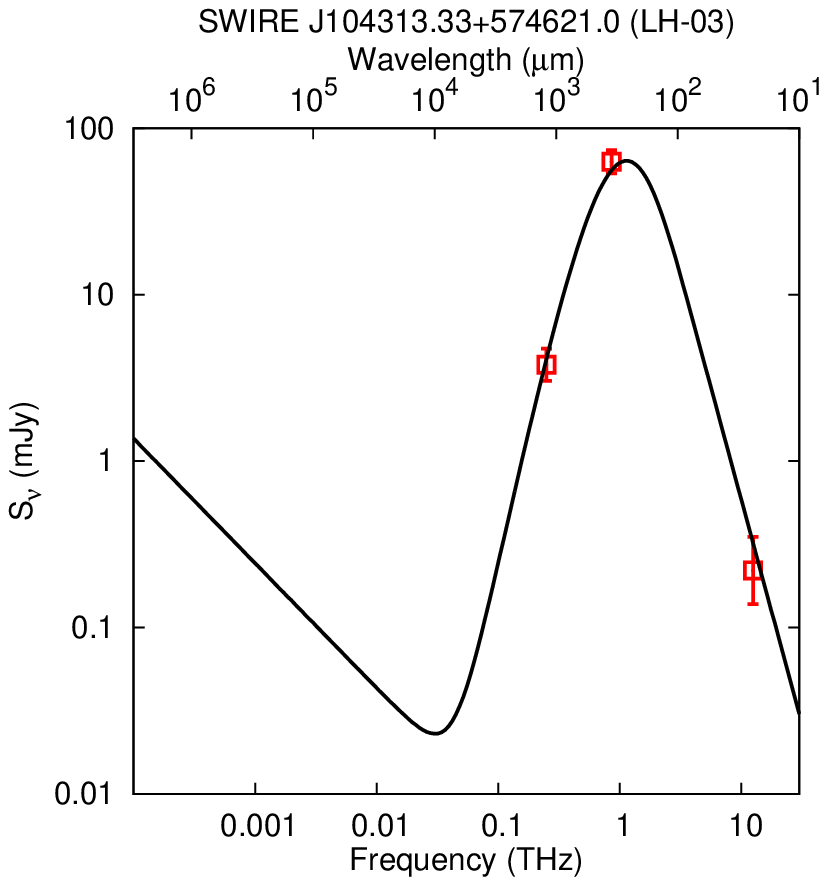}
\includegraphics[width=0.23\textwidth]{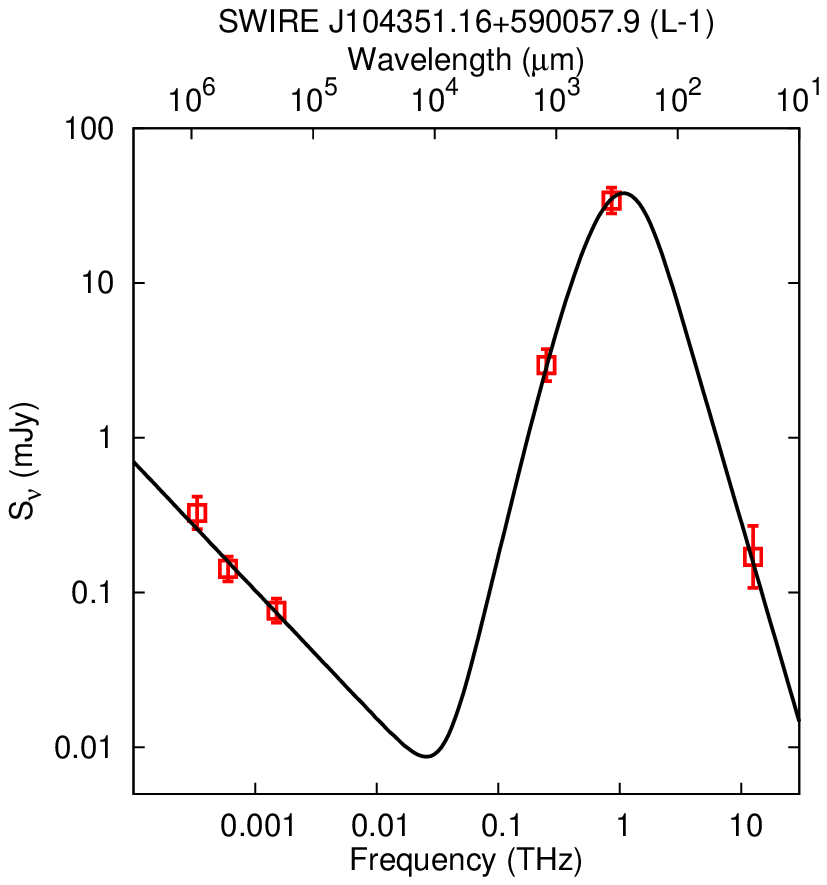} \\
\vspace{6pt}
\includegraphics[width=0.23\textwidth]{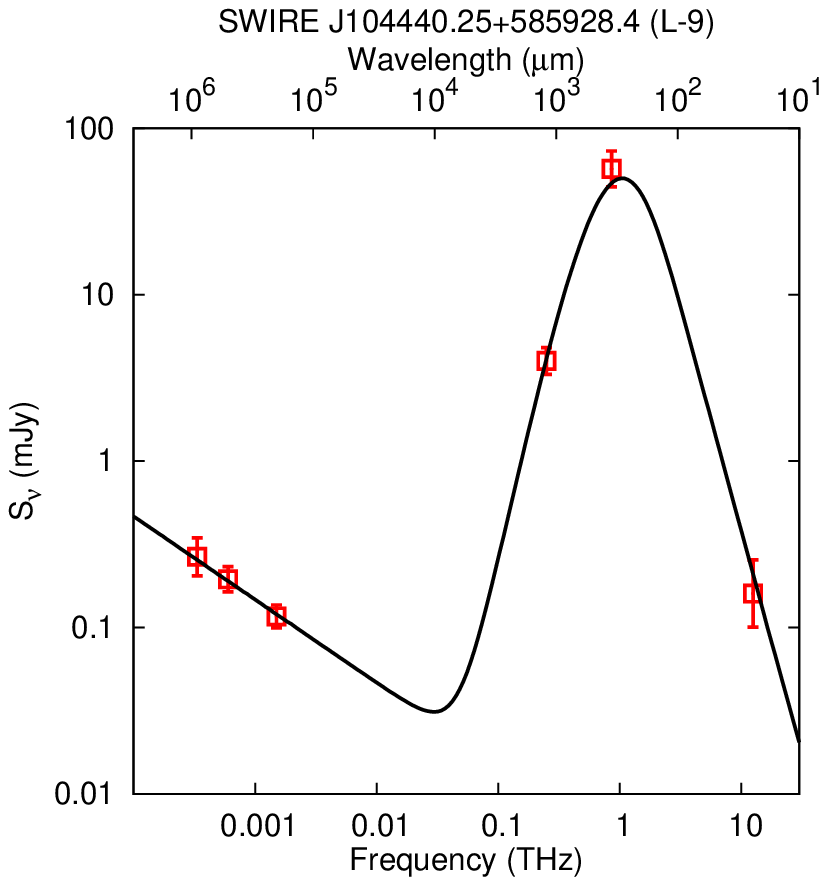}
\includegraphics[width=0.23\textwidth]{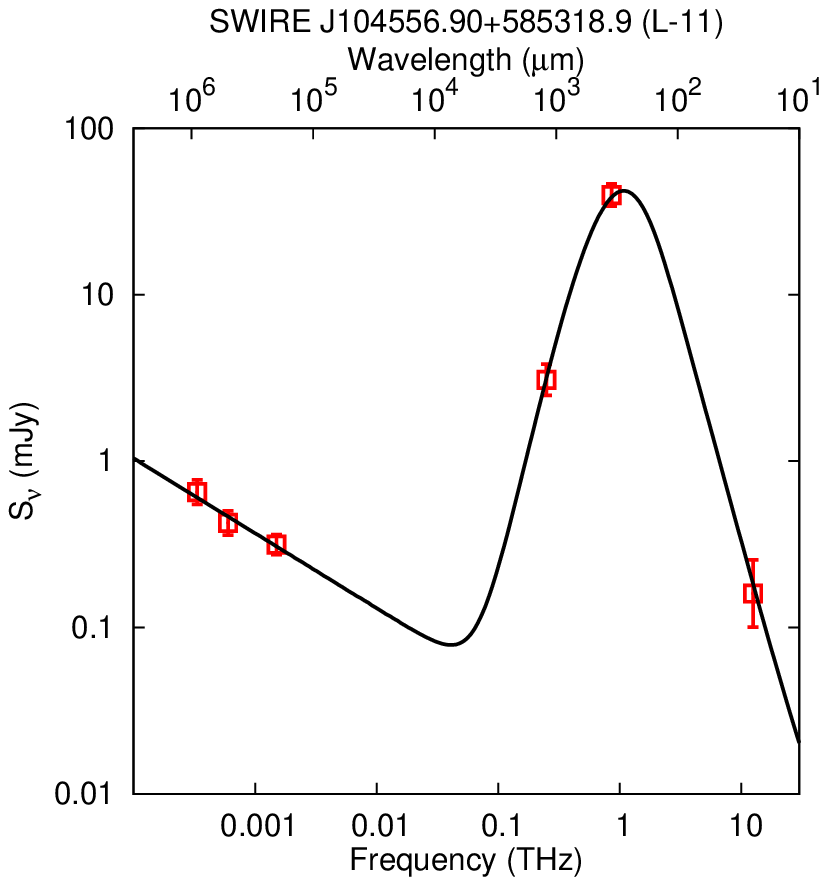}
\includegraphics[width=0.23\textwidth]{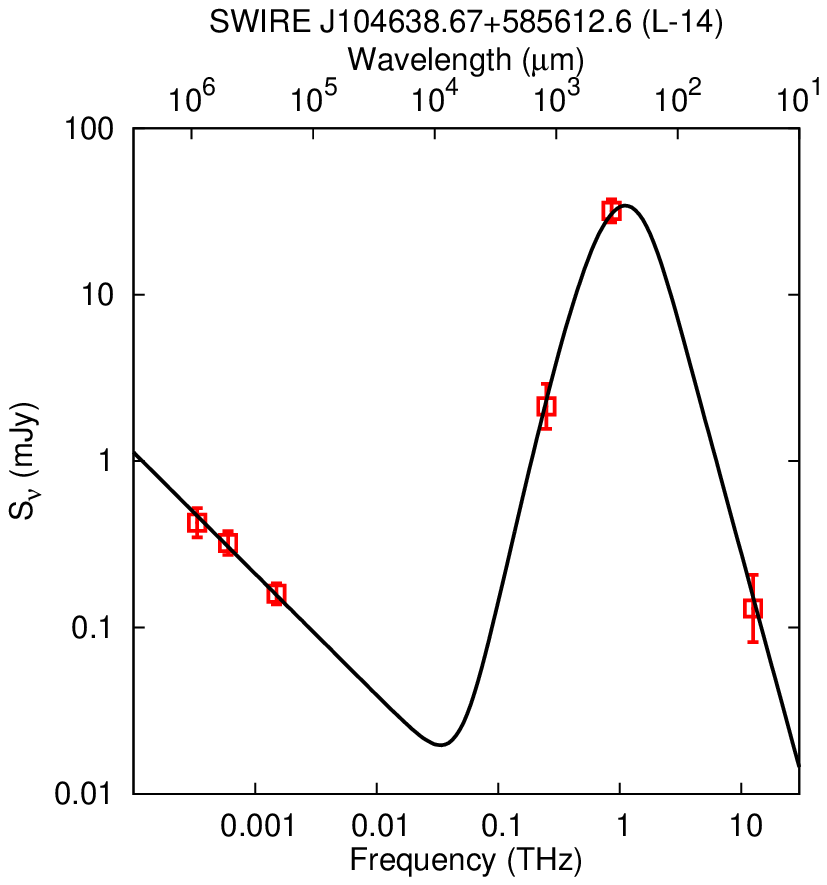}
\includegraphics[width=0.23\textwidth]{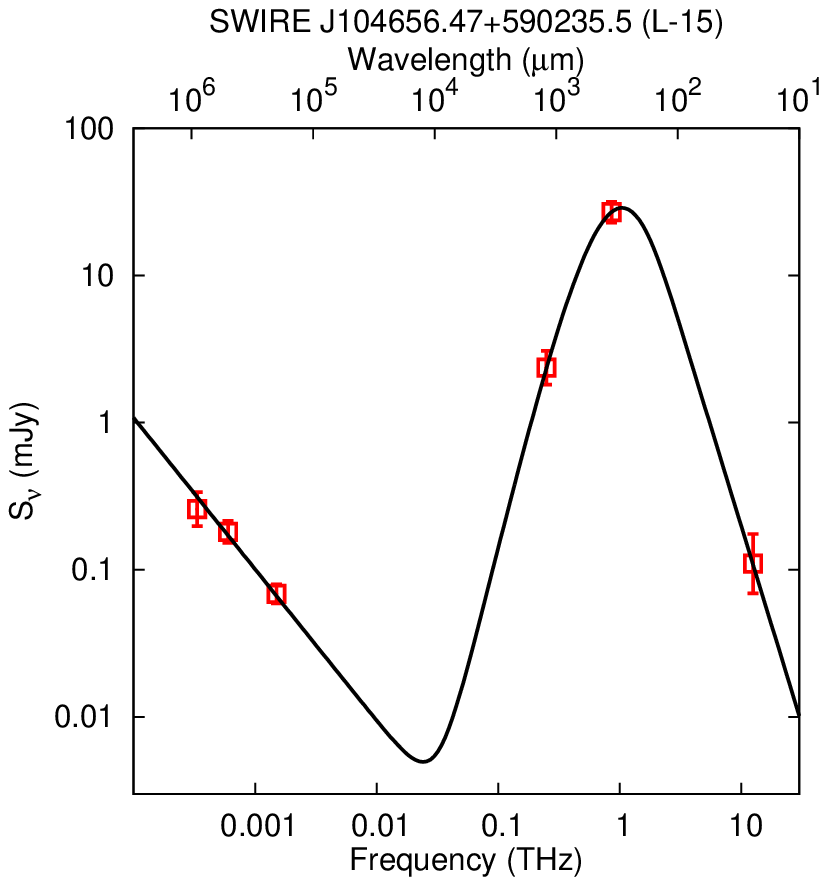} \\
\vspace{6pt}
\includegraphics[width=0.23\textwidth]{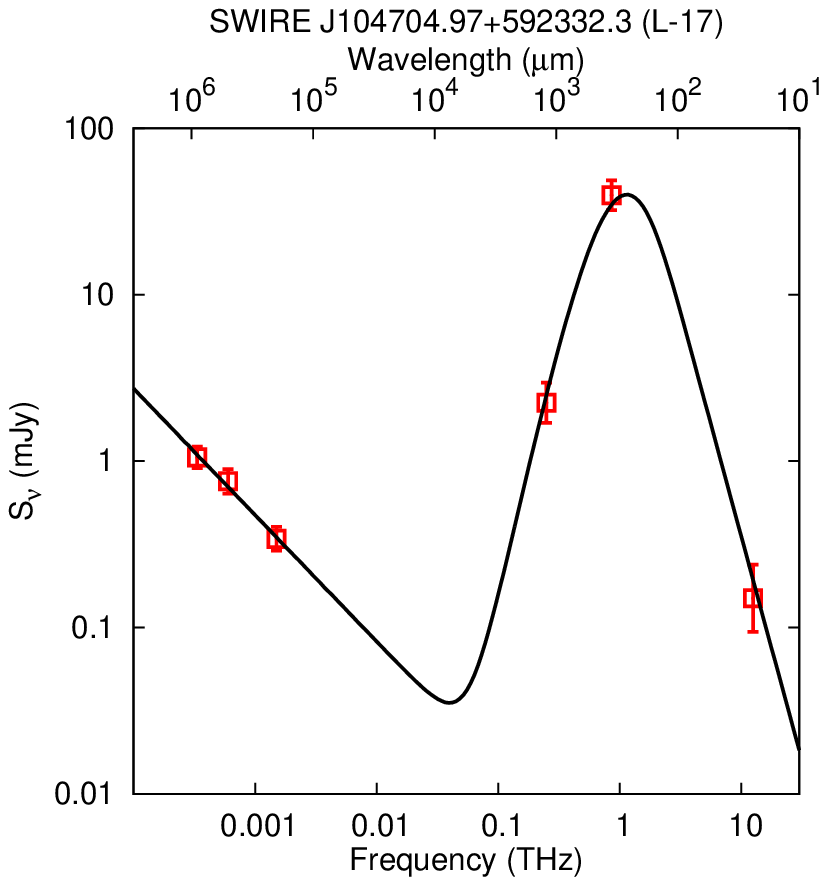}
\includegraphics[width=0.23\textwidth]{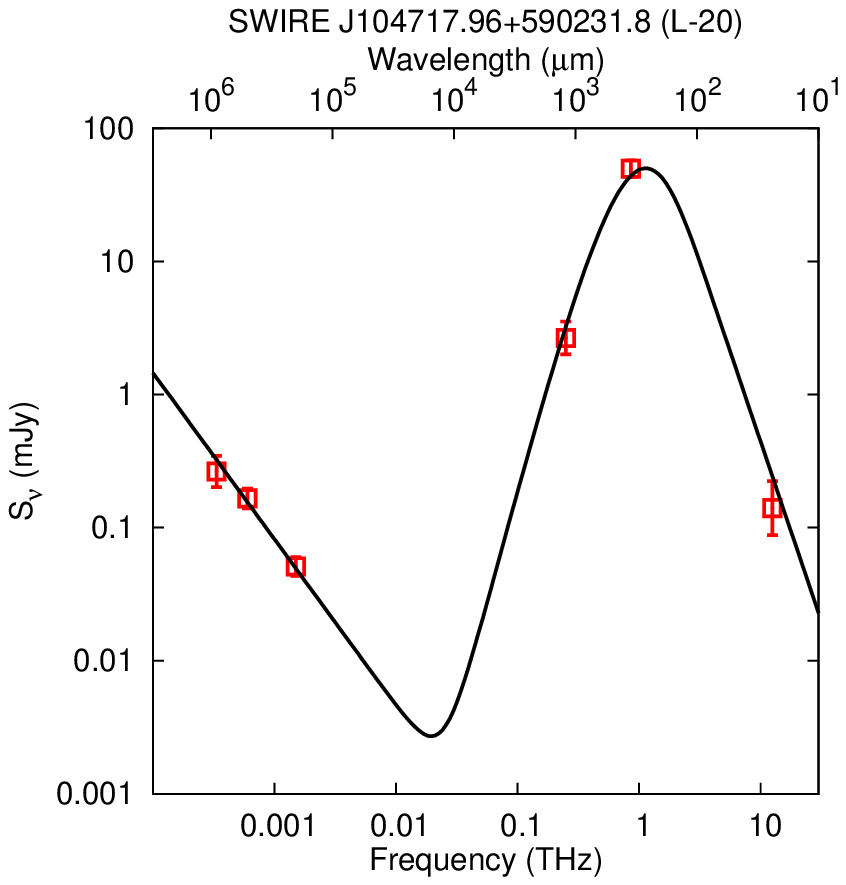}
\includegraphics[width=0.23\textwidth]{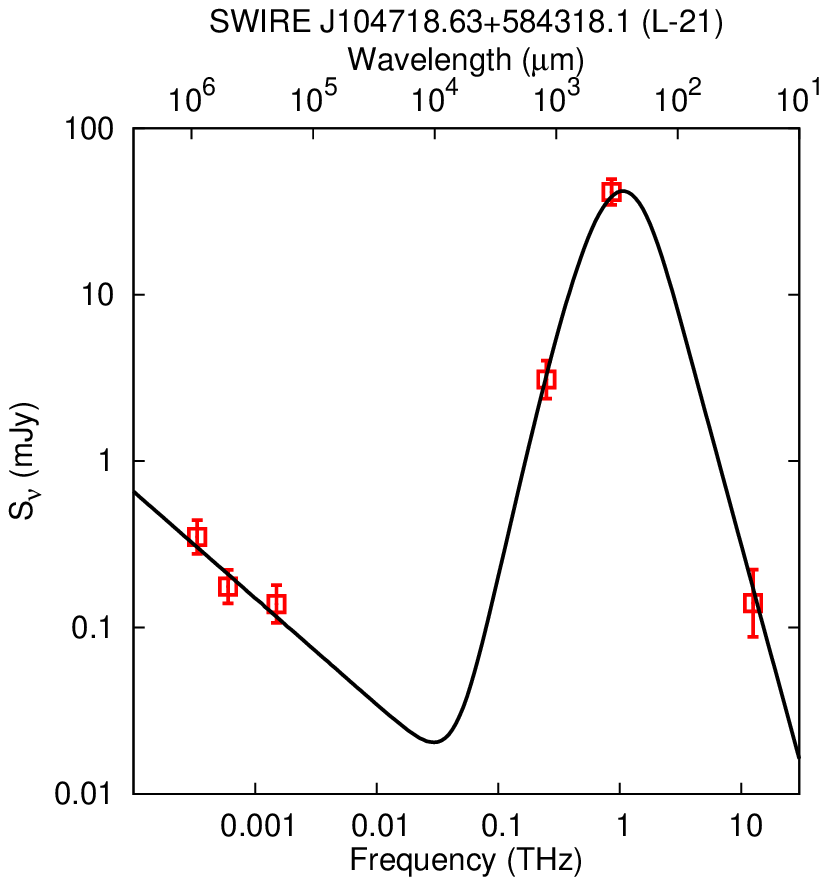}
\includegraphics[width=0.23\textwidth]{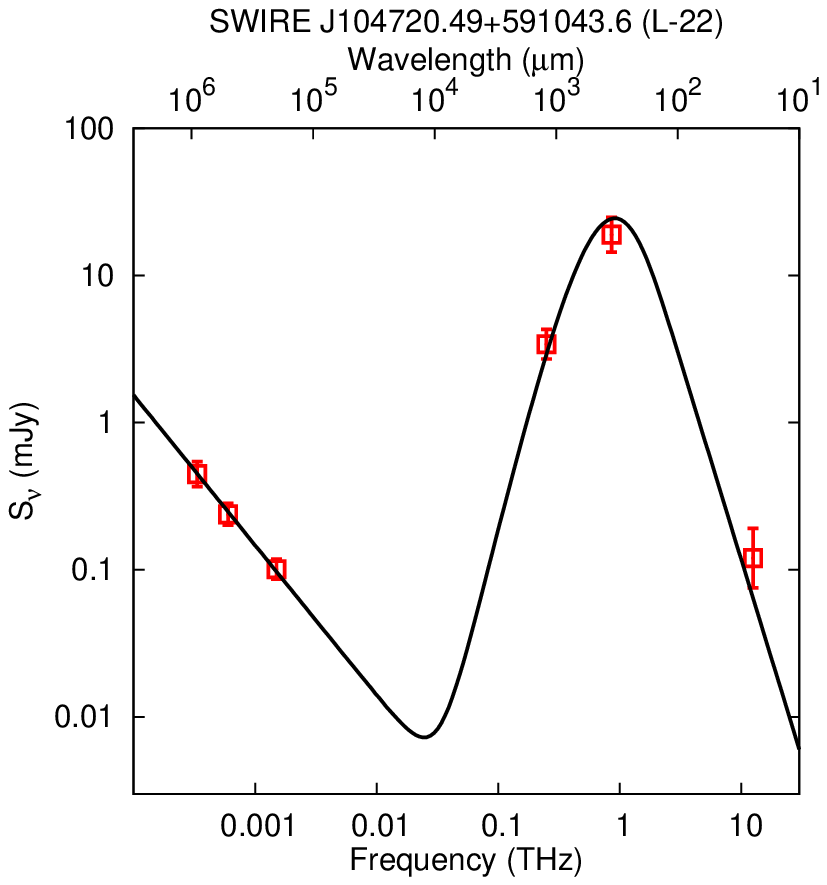} \\
\vspace{6pt}
\includegraphics[width=0.23\textwidth]{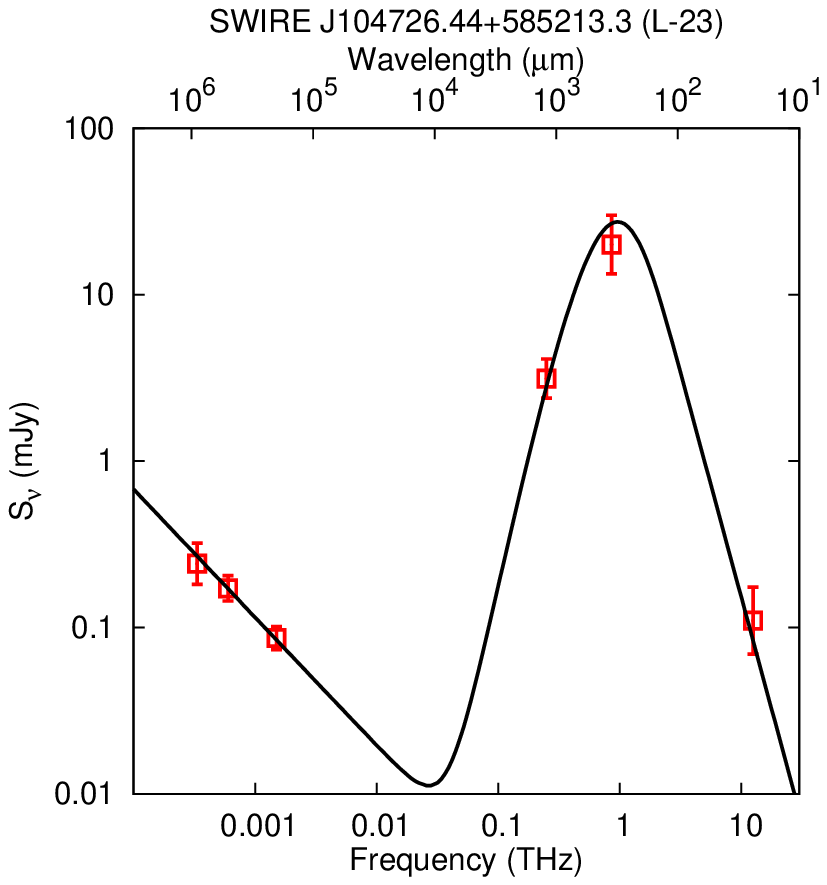}
\includegraphics[width=0.23\textwidth]{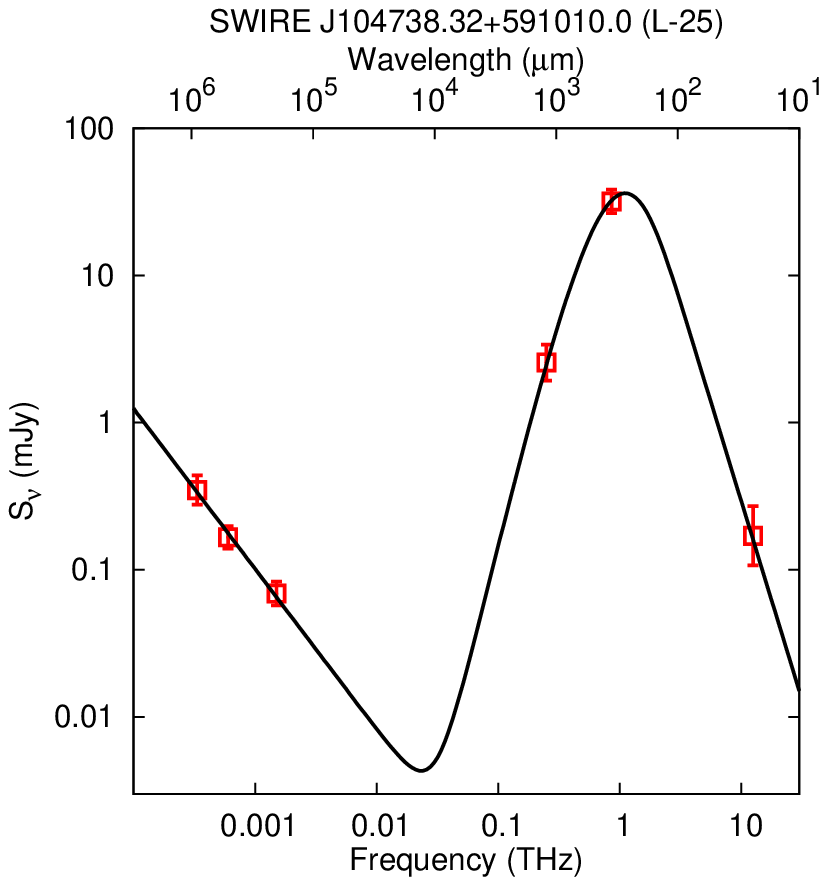}
\includegraphics[width=0.23\textwidth]{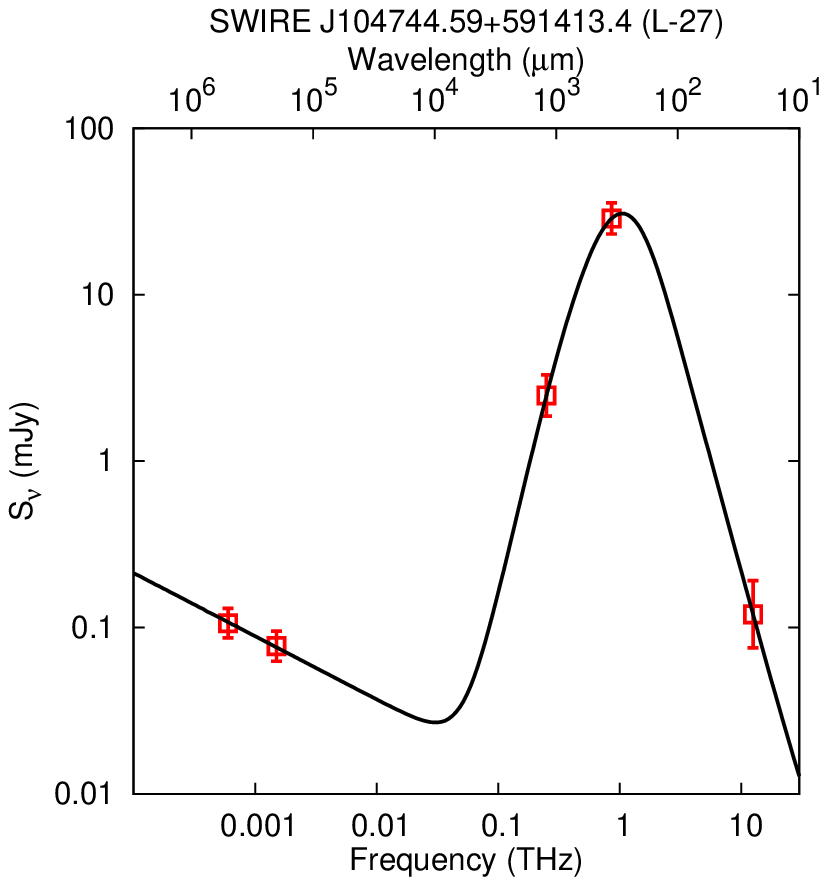}
\includegraphics[width=0.23\textwidth]{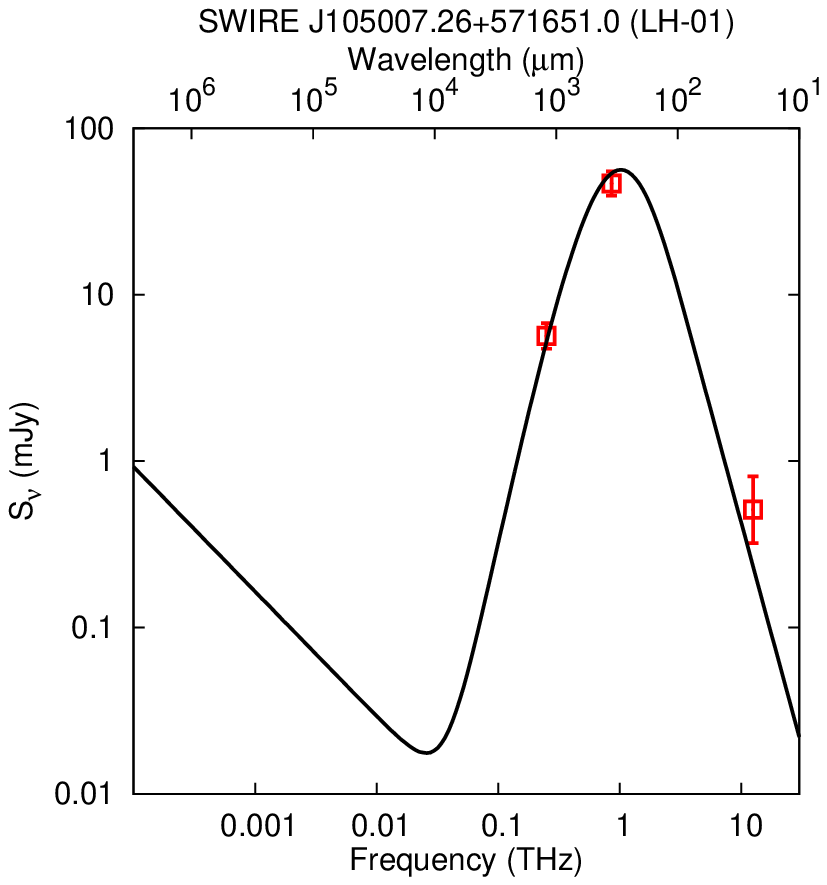} \\
\vspace{6pt}
\includegraphics[width=0.23\textwidth]{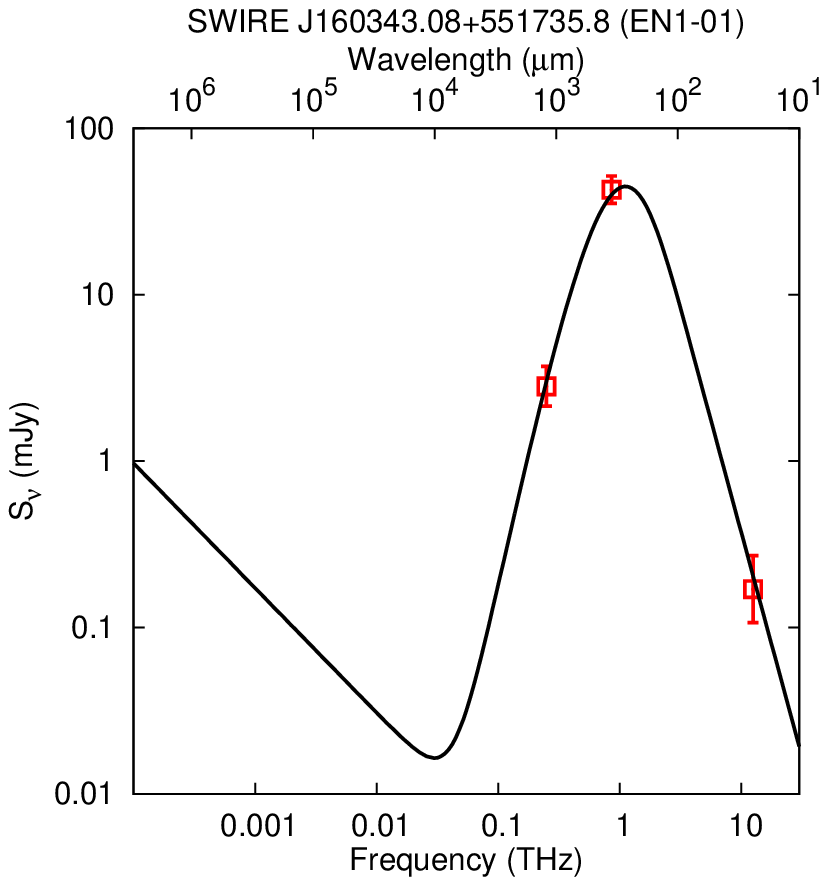}
\includegraphics[width=0.23\textwidth]{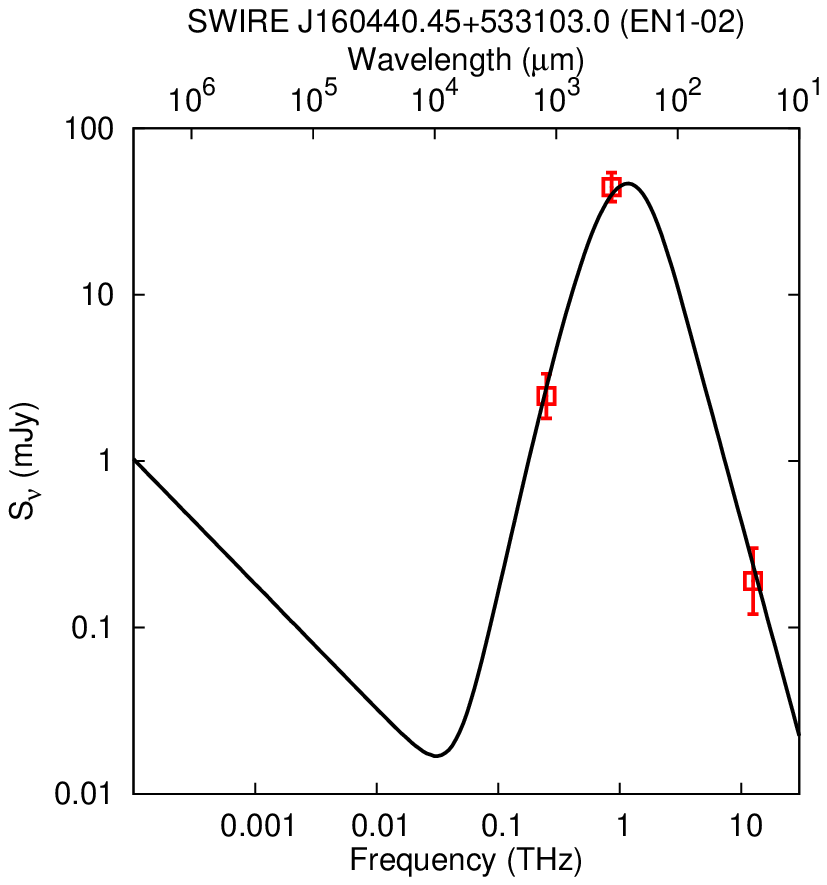}
\includegraphics[width=0.23\textwidth]{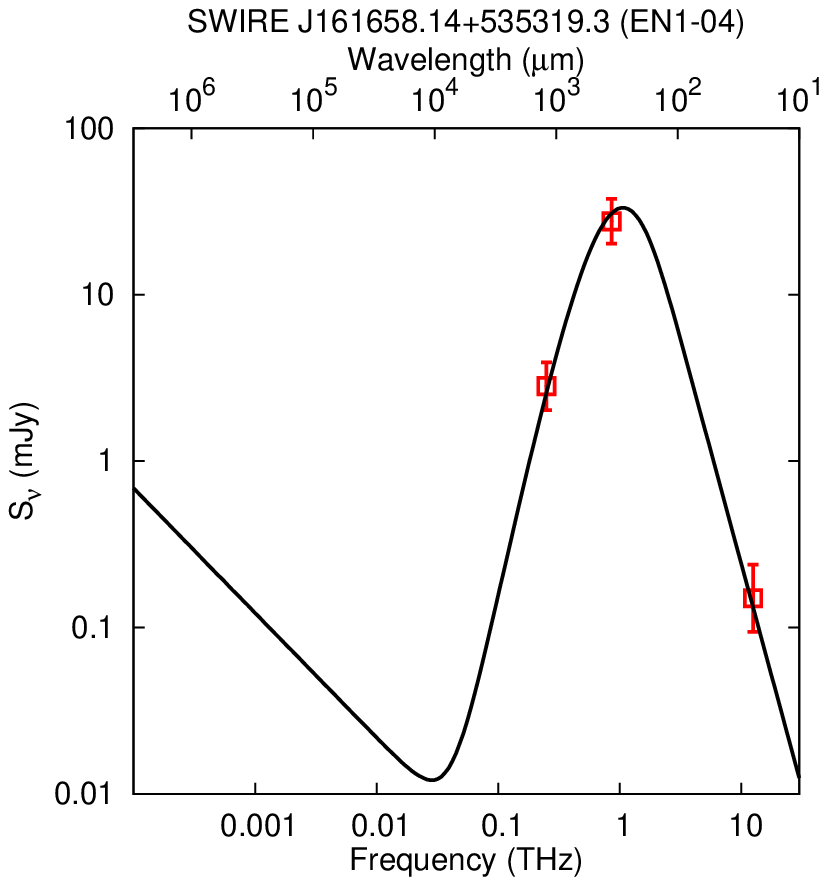}
\includegraphics[width=0.23\textwidth]{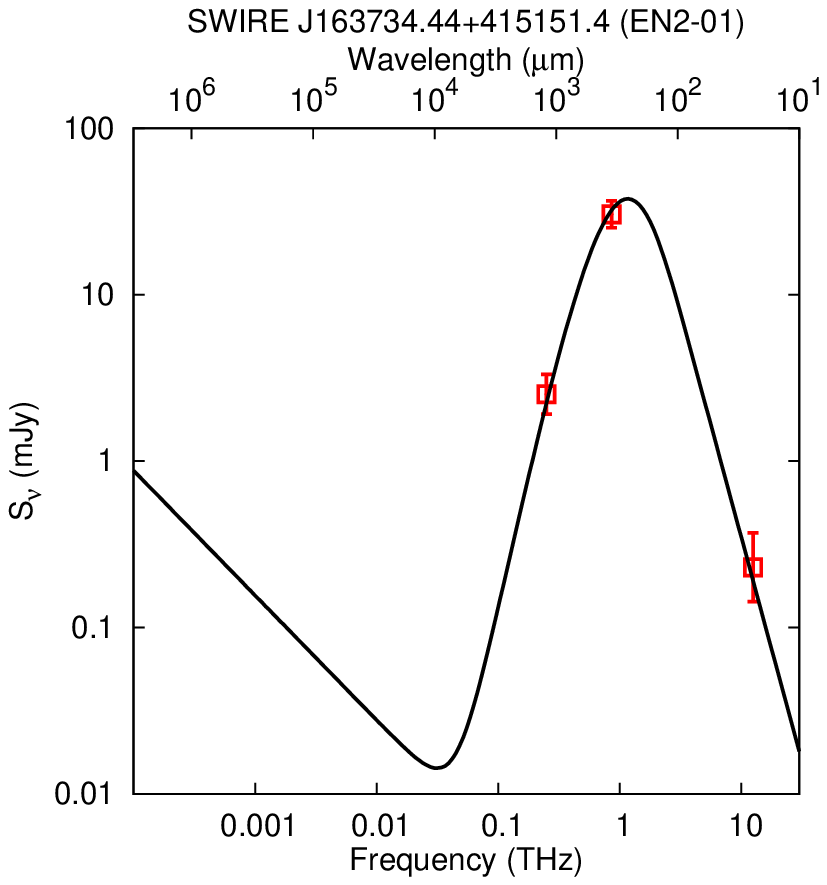} \\
\vspace{6pt}
\caption{Individual power-law temperature distribution SED fits to our sources. The data are shown with hollow red squares and error bars. The 24\um points reflect the expected continuum contribution to the in-band flux densities, according to Figure~\ref{fig:contrib24}, with an uncertainty of 0.25\,dex. The sources without radio data assume $\alpha$=0.75 for the radio spectral index.
} 
\label{fig:swiresed}
\end{figure*}


\begin{thebibliography}{}

\bibitem[Agladze et al.(1996)]{Agladze1996} Agladze, N.~I., Sievers, A.~J., Jones, S.~A., Burlitch, J.~M., \& Beckwith, S.~V.~W.\ 1996, \apj, 462, 1026

\bibitem[Alton et al.(1999)]{Alton1999} Alton, P.~B., Davies, J.~I., \& Bianchi, S.\ 1999, \aap, 343, 51

\bibitem[Appleton et al.(2004)]{Appleton2004} Appleton, P.~N., et al.\ 2004, \apjs, 154, 147

\bibitem[Aravena et al.(2010)]{Aravena2010} Aravena, M., et al.\ 2010, \apj, 708, 36

\bibitem[Austermann et al.(2010)]{Austermann2009} Austermann, J.~E., et al.\ 2010, \mnras, 401, 160


\bibitem[Beelen et al.(2006)]{Beelen2006} Beelen, A., et al.\ 2006, \apj, 642, 694 

\bibitem[Berta(2005)]{Berta2005} Berta, S.\ 2005, PhD thesis, Padua Univ., Italy


\bibitem[Blain et al.(2004)]{Blain2004} Blain, A.~W., Chapman, S.~C., Smail, I., \& Ivison, R.\ 2004, \apj, 611, 725

\bibitem[Caputi et al.(2007)]{Caputi2007} Caputi, K.~I., et al.\ 2007, \apj, 660, 97

\bibitem[Chapman et al.(2003)]{Chapman2003} Chapman, S.~C., Blain, A.~W., Ivison, R.~J., \& Smail, I.\ 2003, \nat, 422, 695 (C03)

\bibitem[Chapman et al.(2005)]{Chapman2005} Chapman, S.~C., Blain, A.~W., Smail, I., \& Ivison, R.\ J.\ 2005, \apj, 622, 772 

\bibitem[Condon \& Broderick(1991)]{CondonBroderick1991} Condon, J.~J., \& Broderick, J.~J.\ 1991, \aj, 102, 1663
\bibitem[Condon(1992)]{Condon1992} Condon, J.~J.\ 1992, \araa, 30, 575

\bibitem[Coppin et al.(2006)]{Coppin2006} Coppin, K., et al.\ 2006, \mnras, 372, 1621

\bibitem[Dale et al.(2001)]{Dale2001} Dale, D.~A., Helou, G., Contursi, A., Silbermann, N.~A., \& Kolhatkar, S.~2001, \apj 549, 215

\bibitem[Dale \& Helou(2002)]{DaleHelou2002} Dale, D.~A., \& Helou, G.~H.\ 2002, \apj, 576, 159

\bibitem[D\'{e}sert, Boulanger \& Puget(1990)]{DBP90} D\'{e}sert, F.-X., Boulanger, F., \& Pouget, J.~L.\ 1990, \aap, 237, 215

\bibitem[Dowell et al.(2003)]{Dowell2003} Dowell, C.~D., et al.\ 2003, Proc.\ SPIE, 4855, 73 


\bibitem[Dunne, Eales \& Edmunds(2003)]{Dunne2003} Dunne, L., Eales, S.~A., \& Edmunds, M.~G.\ 2003, \mnras, 341, 589

\bibitem[Dupac et al.(2003)]{Dupac2003} Dupac, X., et al.\ 2003, \aap, 404, L1


\bibitem[Farrah et al.(2006)]{Farrah2006} Farrah, D., et al.\ 2006, \apj, 641, 17

\bibitem[Farrah et al.(2008)]{Farrah2008} Farrah, D., et al.\ 2008, \apj, 677, 957

\bibitem[F09()]{Fiolet2009} Fiolet, N., et al.\ 2009, \aap, 508, 117

\bibitem[Garrett(2002)]{Garrett2002} Garrett, M.~A.\ 2002, \aap, 384, L19

\bibitem[Greve et al.(2004)]{Greve2004} Greve, T.~R, Ivison, R.~J., Bertoldi, F., Stevens, J.~A., Dunlop, J.~S., Lutz, D., \& Carilli, C.~L.\ 2004, \mnras, 354, 779

\bibitem[Helou et al.(1985)]{Helou1985} Helou, G., Soifer, T., \& Rowan-Robinson, M.\ 1985, \apj, 298, L7
\bibitem[Helou et al.(1988)]{Helou1988} Helou, G., Khan, I.~R., Malek, L., \& Boehmer, L.\ 1988, \apjs, 68, 151

\bibitem[Hildebrand(1983)]{Hildebrand1983} Hildebrand, R.~H.\ 1983, \qjras, 24, 267

\bibitem[Hogg(1999)]{Hogg1999} Hogg, D.~W.\ 1999, arXiv:astro-ph/9905116



\bibitem[Ivison et al.(2007)]{Ivison2007} Ivison, R.~J., et al.\ 2007, \mnras, 380, 199

\bibitem[Ivison et al.(2010)]{Ivison2009} Ivison, R.~J., et al.\ 2010, \mnras, 402, 245

\bibitem[James et al.(2002)]{James2002} James, A., Dunne, L., Eales, S., \& Edmunds, M.~G.\ 2002, \mnras, 335, 753

\bibitem[K\"{o}lbig(1970)]{Kolbig1970} K\"{o}lbig, K.~S.\ 1970, Math. Comput., 24, 679 

\bibitem[Kov\'{a}cs(2006)]{thesis}  Kov\'{a}cs, A.\ 2006, PhD thesis, Caltech

\bibitem[Kov\'{a}cs et al.(2006)]{Kovacs2006} Kov\'{a}cs, A., Chapman, S.~C., Dowell, C.~D., Blain, A.~W., Ivison, R.~J., Smail, I., \& Phillips, T.~G.\ 2006, \apj, 650, 592

\bibitem[Kov\'{a}cs(2008a)]{CRUSH} Kov\'{a}cs, A.\ 2008a, Proc.\ SPIE, 7020, 45

\bibitem[Kov\'{a}cs(2008b)]{scanning} Kov\'{a}cs, A.\ 2008b, Proc.\ SPIE, 7020, 5

\bibitem[Kr\"{u}gel, Steppe \& Chini(1990)]{Krugel1990} Kr\"{u}gel, E., Steppe, H., \& Chini, R.\ 1990, \aap, 229, 17

\bibitem[Leeuw \& Robson(2009)]{Leeuw2009} Leeuw, L.~L., Robson, E.~I.\ 2009, \aj, 137, 517  

\bibitem[Le Floc'h et al.(2005)]{LeFloch2005} Le Floc'h, E., et al.\ 2005, \apj, 632, 169


\bibitem[Leong et al.(2006)]{Leong2006} Leong M.~M., Peng, R., Houde, M., Yoshida, H., Chamberlin, R, \& Phillips, T.~G.\ 2006, Proc.\ SPIE, 6275, 21 

\bibitem[Lonsdale et al.(2003)]{Lonsdale2003} Lonsdale, C.~J., et al.\ 2003, \pasp, 115, 897

\bibitem[L09()]{Lonsdale2009} Lonsdale, C.~J., et al.\ 2009, \apj, 692, 422 (L09)

\bibitem[Owen \& Morrison(2008)]{Owen2008} Owen, F.~N., \& Morrison, G.~E.\ 2008, \aj, 136, 1889

\bibitem[Owen et al.(2009)]{Owen2009} Owen, F.~N., Morrison, G.~E., Klimek, M.~D., \& Greisen, E.~W.\ 2009, \aj, 137, 4846

\bibitem[Polletta et al.(2007)]{Polletta2007} Polletta M., et al.\ 2007, \apj, 663, 81

\bibitem[Pope et al.(2006)]{Pope2006} Pope, A., et al.\ 2006, \mnras, 370, 1185 

\bibitem[Press, Flannery \& Teukolsky(1986)]{Press1986} Press W.\ H., Flannery B.\ P., \& Teukolsky, S.\ A.\ 1986, Numerical Recipes in C: The Art of Scientific Computing (Cambridge: Cambridge Univ.\ Press)

\bibitem[Seaquist et al.(2004)]{Seaquist2004} Seaquist, E., Yao, L., Dunne, L., \& Cameron, H.\ 2004, \mnras, 349, 1428


\bibitem[Sodroski et al.(1997)]{Sodroski1997} Sodroski, T.~J., Odegard, N., Arendt, R.~G., Dwek, E., Weiland, J.~L., Hauser, M.~G., \& Kelsall, T.\ 1997\apj, 480, 173


\bibitem[Tacconi et al.(2006)]{Tacconi2006} Tacconi, L.\ J., et al.\ 2006, \apj, 640, 228


\bibitem[Weedman et al.(2006)]{Weedman2006} Weedman, D., et al.\ 2006, \apj, 653, 101

\bibitem[Weiss et al.(2009)]{Weiss2009} Weiss, A., et al.~2009, \apj, 707, 1201 

\bibitem[Yang(2007)]{Yang2007} Yang, M.\ 2007, PhD thesis, Caltech


\bibitem[Younger et al.(2009)]{Younger2009b} Younger, J.~D., et al.\ 2009, \apj, 704, 803

\bibitem[Yun et al.(2001)]{Yun2001} Yun, M.~S., Reddy, N.~A., \& Condon, J.~J.\ 2001, \apj, 554, 803


\end{thebibliography}
\end{document}